\documentclass[preprint,review,12pt]{elsarticle}

\usepackage{graphicx}
\usepackage{amssymb}
\usepackage{lineno}

\journal{Advances in Space Research}

\begin{document}

\begin{frontmatter}


\title{Performance of NavIC  for studying the ionosphere at an EIA region in India}



\author[a]{Deepthi Ayyagari\corref{c-d54cc1eb1ca4}}
\ead{nagavijayadeepthi@gmail.com}\cortext[c-d54cc1eb1ca4]{Corresponding author.}
\author[a]{Sumanjit Chakraborty}
\ead{sumanjit11@gmail.com}
\author[a]{Saurabh Das}
\ead{das.saurabh01@gmail.com}
\author[b]{Ashish Shukla}
\ead{ashishs@sac.isro.gov.in }
\author[c]{Ashik Paul}
\ead{ap.rpe@caluniv.ac.in}
\author[a]{Abhirup Datta}
\ead{abhirup.datta@iiti.ac.in}

\address[a]{Discipline of Astronomy, Astrophysics and Space Engineering\unskip, IIT Indore\unskip, Simrol \unskip, Indore\unskip, 453552, Madhya Pradesh, India}
  	
\address[b]{Space Applications Centre\unskip,Indian Space Research Organization, Ahmedabad 380015,Gujarat,India}

\address[c]{Institute of Radio Physics and Electronics, University of Calcutta, Kolkata 700009, West Bengal, India.}

\begin{abstract}

This paper emphasizes on  NavIC's performance in ionospheric studies over the Indian subcontinent region. The study is performed using data of one year (2017-18) at IIT Indore, a location near the northern crest of Equatorial Ionization Anomaly (EIA). It has been observed that even without the individual error corrections, the results are within $\pm20\%$  of NavIC VTEC estimates observed over the 1\ensuremath{^{\circ}} x 1\ensuremath{^{\circ}} grid of IPP surrounding the GPS VTEC estimates for most of the time.. Additionally, ionospheric response during two distinct geomagnetic storms (September 08 and 28, 2017) at the same location and other IGS stations covering the Indian subcontinent using both GPS and NavIC has also been presented. This analysis revealed similar variations in TEC during the geomagnetic storms of September 2017, indicating the suitability of NavIC to study space weather events along with the ionospheric studies over the Indian subcontinent.

\end{abstract}

\begin{keyword}
Ionosphere; NavIC; GPS; TEC; Iono-delay; Geomagnetic Storms
\end{keyword}
\end{frontmatter}


\section{Introduction}
\label{S:1}

The ionosphere has been studied for many decades using the Faraday rotation effect on a linear polarized propagating plane wave. However, global coverage of the Global Positioning System (GPS) (\citet{dpt:1}) the advent to have ionospheric measurements at multiple points on the ionosphere from a single location (\citet{dpt:15}). Despite its global coverage and improvised technical applications, the availability of GPS satellites at any time instant from any location is limited to 6-8 satellites, implying 6-8 sampling points on the ionosphere. The availability of other global as well as regional satellite systems such as the Navigation with Indian Constellation (NavIC) becomes useful for ionospheric research as it increases the number of ray paths through the ionosphere. The ionosphere, as well as space weather, play a  major role in areas such as satellite communication, remote sensing and, electrical systems. GPS-derived TEC data have been extensively used for ionospheric studies and to analyze and validate both ionospheric models and space weather monitoring applications (\citet{dpt:16}). 

NavIC is a regional satellite navigation system with a combination of three Geostationary Earth Orbit (GEO) and three Geosynchronous Orbit (GSO) in its space segment and is developed by the Indian Space Research Organization (ISRO). It is designed and developed to provide positional accuracy information to the Indian users and also extends to the region of 1500 km from its boundary, designated to be its primary service area. NavIC has a provision for an extended service area that lies between the primary service area and area enclosed by the rectangular grid from 30\ensuremath{^{o}}S to 50\ensuremath{^{o}}N in latitude to 30\ensuremath{^{o}}E to 130\ensuremath{^{o}}E in longitude.  Furthermore, these satellites broadcast signals in 24MHz bandwidth of spectrum in the L5 and S1 band with carrier frequencies of 1176.45 MHZ and 2492.028 MHz respectively (\citet{dpt:3}). 

Recent studies using NavIC data, reveal the signal strength and quality of NavIC satellites to be reliable and can be used for studies of the iono-delay range calculations aiding in TEC estimation (\citet{dpt:13}). Moreover, a method that involves ionospheric gradient analysis using a weighted least square algorithm confirms the increments of VTEC accordingly with the effects of geomagnetic storm (\citet{dpt:18}). \citet{dpt:19} did a comparative study of GPS-TEC from the IGS station at IISC, Bengaluru and International Reference Ionosphere (IRI) derived TEC with NavIC estimated TEC based on iono-delay measurements as well as pseudo-ranges, and found it to be in agreement with the model estimated TEC, thus concluding NavIC signals to be not only reliable for upper atmospheric sounding but also for the navigational applications. Furthermore (\citet{dpt:20}) have presented in their study on how to determine VTEC using a dual-frequency method along with a short term analysis on the diurnal variation of NavIC data. They found the fourth-order polynomial VTEC values in agreement with the Klobuchar single frequency estimation (\citet{dpt:21}) of VTEC using a cosine angle mapping function. \citet{dpt:22} developed a novel algorithm that estimates the ionospheric delay and provides ionospheric corrections, during depleted ionospheric conditions, using single frequency (L5/S1) data from the IRNSS receiver.

The Ionosphere has been studied extensively by several researchers over the past few decades using GPS data but for the first time in this paper, to the best of our knowledge, ionosphere and geomagnetic storm effects on it have been monitored using a combination of both NavIC and GPS data. It is to be noted here that the TEC estimation process is an involved process due to the presence of several biases and error sources which is not the focus of this work. These biases (code and carrier phase error/bias estimates) and the corrections involved in estimating the measurements have been open research in GPS even today. There has not been a definite standard methodology to mitigate these error sources and biases as per the available literature (\citet{dpt:29},\citet{dpt:33},\citet{dpt:36},\citet{dpt:38},\citet{dpt:39}, \citet{dpt:40}). In this respect, NavIC is a regional system which is still unexplored in terms of ionospheric error sources, mitigation of these biases (\citet{dpt:37}) and assessment of the performance of different navigation satellite systems for the estimation of TEC.

\section{Data}

\label{S:3}

The Discipline of Astronomy, Astrophysics and Space Engineering of Indian Institute of Technology (DAASE), Indore (Lat:22.52\ensuremath{^{o}}N, Lon:75.92\ensuremath{^{o}}E; Magnetic dip: 32.23\ensuremath{^{o}}N) operates a multi-constellation, multi-frequency GNSS receiver since May 2016 capable of receiving signals, GPS (L1, L2 and L5), GLONASS (G1, G2 and G3) and GALILEO (E1,E5, E5a, E5b,E6). A NavIC receiver, provided by Space Applications Centre, ISRO, capable of receiving GPS L1, NavIC L5 and S1 signals, was operational since May 2017. The present study also utilizes data from the International GNSS Service (IGS) (\citet{dpt:4}) for the three stations: Lucknow (Lat:26.91\ensuremath{^{o}}N, Lon:80.95\ensuremath{^{o}}E; Magnetic dip: 39.75\ensuremath{^{o}}N), Hyderabad (Lat:17.41\ensuremath{^{o}}N, Lon:78.55\ensuremath{^{o}}E; Magnetic dip: 21.69\ensuremath{^{o}}N) and Bengaluru (Lat:13.02\ensuremath{^{o}}N, Lon:77.57\ensuremath{^{o}}E ; Magnetic dip: 11.78\ensuremath{^{o}}N). The GPS and IGS data have been analyzed by masking the elevation angle $\geq$20\ensuremath{^{o }} to reduce multipath error.

\section{Methodology}

The frequency dependence of the ionospheric effect of TEC mentioned by several researchers (\citet{dpt:29}, \citet{dpt:2}, \citet{dpt:24}, \citet{dpt:23,dpt:28}, \citet{dpt:25}, \citet{dpt:27}, \citet{dpt:36}) is given as:
\begin{equation}
\rho_{iono,f} = 40.3 .\frac{STEC}{f^2}
\end{equation} 
where \textit{$\rho_{iono}$} is the iono-delay in m, \textit{f} is the operational frequency of the signal emitted by satellites in Hz. The iono-delay is approximated up to first order of the ionospheric term. 

The delay term mentioned in equation (1) appears in the standard estimation process of pseudo-range measurement equations (for the NavIC receiver(\citet{dpt:30}) and GNSS receiver (\citet{dpt:31}) used in the present study) at f1 and f2 frequencies (S1, L5 for NavIC and L1, L2 for GPS), composed of true range $r$, receiver and satellite clock biases $\Delta t_r$, $\Delta t_s$ respectively, ephemeris error component $e$ along the satellite-receiver line-of-sight, ionospheric delay $\rho_{iono,f}$, tropospheric delay $T_t$, satellite and receiver instrument delays $k^s$ and $k^r$, code phase multipath error $M_p$ and random noise $\nu$ are given by
\begin{equation}
\tilde{P}_{f1} = r +  c (\Delta t_r - \Delta t_s) + e + \rho_{iono,f1} + T_t + ck_{f1}^r - ck_{f1}^s + M_{pf1} + \nu_{f1}
\end{equation} 
\begin{equation}
\tilde{P}_{f2} = r +  c (\Delta t_r - \Delta t_s) + e + \rho_{iono,f2} + T_t + ck_{f2}^r - ck_{f2}^s + M_{pf2} + \nu_{f2}
\end{equation} 

For the dual-frequency users the difference between  equations (2) and (3) is used so as remove the satellite instrument delays through the satellite clock corrections by forming the iono-free pseudo-ranges which is given as:
\begin{equation}
  {\rho}_{IF} = \frac{\rho_{iono,f1} - (\gamma) {\rho}_{iono,f2}}{1 -(\gamma)} 
\end{equation}
where $\gamma = (\frac{F_{f2}}{F_{f1}})^2$. The equation (4) is further simplified based on the values of f1 and f2 which are vary with NavIC ($\gamma$ = 0.2229) and GPS ($\gamma$ = 1.6469) signal-carrier frequencies in order to obtain the TEC estimates.

To measure TEC along the line of sight, a simplified model which assumes the ionosphere as a thin, uniform-density shell around the Earth, located near the mean altitude ($h_{I}$ $\sim $ 350 km) of maximum TEC is considered (\citet{dpt:5},\citet{dpt:26}). Using spherical geometry, a slant intersection with this shell model can be determined and a VTEC can be inferred using appropriate mapping function. The intersection between the line-of-sight and this shell is called the Ionospheric Pierce Point (IPP). The perpendicular projection of this point onto the earth's surface is called the sub-ionospheric point. At any point azimuth (\textit{Az}) and elevation (\textit{E}) of the line-of-sight vector from user to satellite along with user's latitude-longitude ($\phi_u,\lambda_u $) is necessary to calculate the IPP ($\phi_{pp},\lambda_{pp} $) locations and is given as:
\begin{equation}
\psi_{pp}= \frac{\pi}{2}-E-sin^{-1}{\Big[\frac{{R_e} . cos{(E)}}{R_e{+h_I}}\Big]}\ 
\end{equation}

\begin{equation} 
\phi_{pp}=\ sin^{-1}\ \big[sin\phi_{u}*cos\psi_{pp} + cos\phi_{u} * sin\psi_{pp}* cos(Az)\big]
\end{equation}

\begin{equation} 
\lambda_{pp} = \lambda_{u}+sin^{-1}\bigg[\frac {sin\psi_{pp}\ast sin(Az)}{cos\phi_{pp}}\bigg]
\end{equation}

Using equations (2), (3) and (4) the IPP latitude-longitude values from the user's latitude-longitude positions have been calculated. The IPP's for both NavIC and GPS is shown in Fig.1 for a typical day. This figure shows the map of the IPP latitude spread of NavIC GEO between 20$^o$N-22$^o$N, GSO between 17$^o$N-24$^o$N, IPP longitude spread for both GEO and GSO between 71$^o$E-83$^o$E while the spread of GPS IPP latitude is 15$^o$N-28$^o$N and longitude is 68$^o$E-84$^o$E. The interaction of trans-ionospheric radio waves with the ionospheric plasma causes a first-order propagation delay which is proportional to the inverse of the squared radio frequency (\textit{1/f}\ensuremath{^{2}}) and the integrated electron density (TEC) along the ray path. Hence, TEC describing the first-order ionospheric range error is of particular interest in GNSS applications (\cite{dpt:6}). The STEC is defined as the integral of the electron density $n_e$ along the ray path $s$ between a satellite $S$ and a receiver $R$ and given by:

\begin{equation}
TEC = \int_{S}^{R}  {n_e} . {ds} 
\end{equation}

Due to the dispersive nature of the ionosphere in L and S-band frequencies, the STEC may be derived from dual-frequency GNSS measurements. The STEC is then converted to VTEC by applying a mapping function. Assuming a single layer spherical ionosphere, the corresponding mapping function converts STEC obtained from equation (1) to VTEC and vice versa (\citet{dpt:23}, \citet{dpt:7}, \citet{dpt:36}, \citet{dpt:20}):
\begin{equation}
           M(E)  = \Bigg[ \bigg[ 1 -  \Big[\frac{{R_e} . cos{(E)}}{R_e+h_I}\Big]^{2}\bigg]\Bigg]^{-1/2}
\end{equation}

Here {$R_e$} is the radius of the Earth (6371 km), {$h_I$} denotes the altitude of the thin shell model of the ionosphere (350 km)  and ({$E$}) is the elevation angle of the space vehicle.

\section{Present Study}

\subsection{Comparative study of GPS and NavIC}

Based on IPP locations of NavIC satellites at every time instant, the IPP values of GPS are considered in a grid of 1\ensuremath{^{\circ}} x 1\ensuremath{^{\circ}} surrounding the NavIC ray path. Hence, for each time instant, corresponding to a TEC value estimated by NavIC, few TEC values estimated from GPS satellites have been observed. Here the assumption is that ionosphere is invariant (\cite{dpt:10}) in a 1\ensuremath{^{\circ}} x 1\ensuremath{^{\circ}} grid and the same ionosphere is sampled by NavIC and GPS. Of course, this assumption may not be always true in a strict sense; however, reducing the grid size further for the present purpose will be impractical due to the limited availability of GPS ray paths over any location. The VTEC values estimated by these two navigation systems, are then compared for a period of one year, starting from September 1, 2017, to September 30, 2018. The number of satellites available in this period in the invariant ionosphere within this 1\ensuremath{^{\circ}} x 1\ensuremath{^{\circ}} grid for these navigation systems are shown in Fig.2 which clearly shows the unavailability of GPS ray paths under NavIC satellite PRN-5. This indicates that NavIC gives a better spatial as well as temporal coverage over this 1\ensuremath{^{\circ}} x 1\ensuremath{^{\circ}} grid invariant ionosphere, especially for this location.

To get a broader view on the process of estimation of VTEC from these constellations of satellites under varied ionospheric conditions, the total period is divided into two parts, namely quiet and disturbed days based on the $K_{p}$ index. $K_{p}$ index indicates the disturbances in the horizontal component of the earth's magnetic field in the range from 0{\textendash}9. $K_{p}$ index value up to 4 signifies a calm period and 5 or more indicates a geomagnetic storm (\citet{dpt:8}).

After synchronizing the VTEC estimates from both the receivers based on the same instant of time in 1\ensuremath{^{\circ}} x 1\ensuremath{^{\circ}} grid, the NavIC VTEC are plotted against GPS VTEC estimates. These are shown as Fig.3(a),4(a) and 5(a) each representing the total period, quiet period and disturbed period respectively. It's a clear observation that these plots show a linear relationship between NavIC and GPS VTEC values. To compare the values of VTEC of NavIC and GPS based for the analysis the values have been brought to the same reference level based on the quiet time ionosphere (\citet{dpt:32},\citet{dpt:33},\citet{dpt:34},\citet{dpt:35}) which was not more than 5 TECU for each day of analysis and thus obtained plots for NavIC VTEC to GPS VTEC are represented in Fig.3(b),4(b) and 5(b), for whole period, quiet and disturbed period, respectively. Later, on investigating based on the pattern shown in  Fig.3(b), 4(b) and 5(b), it has been discovered that NavIC data has some anomalous VTEC values as shown in Fig 6(a-b) especially during the quiet ionospheric time. The anomalous data was spotted for 167 days during the whole period of analysis. This anomalous data is because of zero/negative values in one of the pseudorange measurements of the NavIC satellite range estimates. So the data has been corrected using the same reference level of VTEC values of NavIC and GPS neglecting all the anomalous values. After completely removing the anomalous data for the period of analysis then the scatter plots thus obtained for NavIC VTEC to GPS VTEC are represented in Fig. 3(c),4(c) and 5(c), for the whole period, quiet and disturbed period, respectively. The data of VTEC values w.r.t to before and after the anomalous data detection, reduced so as to bring it to the quiet time ionospheric value is represented in Fig. 7(a-b) as the overestimated values for each of the NavIC satellites.

Even after correcting for these anomalous values, from each of the PRN's data set, there remains some difference in GPS to NavIC measured VTEC. For a simple understanding, the difference in TEC estimates of NavIC to GPS is denoted as $\Delta TEC_{NG}$. The $\Delta TEC_{NG}$ values are positive for more than 65\% of the data. The spread of data points around the mean line maybe because the ionosphere can change over short distances as well as in the invariant one-degree grid. The spread of the data points in disturbed days are found to be more than the calm days and support this fact. Nonetheless, these plots show that the NavIC constellation's observables are as consistent as GPS observed TEC. The $\Delta TEC_{NG}$ distribution for the anomalous corrected as well as for the uncorrected values for a total period of one year, quiet and disturbed days respectively are shown in Fig.8(A-B-C),9(A-B-C) and 10(A-B-C). In each of these figures the plots (A) represents the uncorrected value's $\Delta TEC_{NG}$ distribution for each of the satellite vehicles of NavIC and plots from NavIC (PRN-2 to PRN-7) in ascending order, (B) likewise represents the reference level corrected values $\Delta TEC_{NG}$ distribution for each of the satellite vehicle of NavIC including the anomalous data and (C) represents the $\Delta TEC_{NG}$ distribution of each of the NavIC PRN for the anomalous corrected values after removal of the anomalous data respectively. The peak values in the $\Delta TEC_{NG}$ distribution plots (Fig.8-10(A-B-C), during the period of one year of this analysis, remained between -20\% to 20\% in both the cases (b-c).

In this case, one can infer that the difference in magnitude of NavIC estimated VTEC and the GPS estimated VTEC maybe because of the altitude difference between the two constellations. However, the data values show a similar kind of distribution during disturbed days when compared to all and quiet days as a result of a greater number of the data points. After removing the anomalous data represented in both cases as (c) it was observed that the $\Delta TEC_{NG}$ distribution is almost symmetrical which indicates that NavIC estimated the VTEC is equal to GPS estimated VTEC.

The plots in Fig. 11(a-b-c), 12(a-b-c) and 13(a-b-c) show the whole $\Delta TEC_{NG}$ distribution of the all the values of the NavIC Constellation for the same period of one year, quiet days and disturbed days for the uncorrected (a), corrected based on reference level value mention above (b) and after removing the anomalous data (c). In all the Fig.11-12-13 the case (c) reveals symmetric distribution, where the peak values for all the three distributions are between -20\% to 20\%. Such a distribution concludes the consistent behavior of NavIC. This equally proves the reliability of the performance of NavIC estimates in the invariant ionospheric grid. The more symmetrical nature of Fig 13(c) when compared to Fig 11(c) and 12(c) is due to the less number of data points for disturbed days during the period of analysis.

\subsection{Ionospheric response to geomagnetic storms}

Geomagnetic storms are disturbances of the Earth's magnetosphere and are caused by a solar wind shock waves which strike the Earth's magnetic field about one to two days after the event. They are associated with CME, Co-rotating Interaction Region (CIR) and solar flares (\citet{dpt:9}). A vital parameter in identifying severity of geomagnetic storms is the Disturbance storm time (Dst) index, (\citet{dpt:11}) which measures the horizontal component of the Earth's magnetic field (H) in nano Tesla ({\textit{nT}}). During such disturbances, this field gets depressed and its magnitude, which is axially symmetric in nature, varies with the storm time or the time measured from the onset of a storm. Severity of geomagnetic storms can be classified as moderate storm (-50 nT $\leq$ Dst $<$ -100 nT) and intense storm (-100 nT $\leq$ Dst $<$ -200 nT) (\citet{dpt:11}). Most recently, (\citet{dpt:41}) have studied the influence of CME followed by CIR induced intense storms of October 2016 and the CME induced storm of May 2017 over the low-latitude ionization of the Indian subcontinent, thus bringing forward the importance of investigating the ionospheric effects of space weather over such a dynamic region. The present study includes an intense storm on September 8, 2017, where Dst reached a minimum of -124 nT and a moderate storm on September 28, 2017, where Dst reached a minimum -55 nT as shown in Fig. 14. It is to be noted that (\citet{dpt:14}) studied the impact of the intense storm September 8, 2017, on NavIC over five stations of the Indian subcontinent, whereas in this paper, two distinct storms of September 2017 are studied with both NavIC and GPS VTEC data.

\subsubsection{Storm of September 5-9, 2017}

According to the National Oceanic and Atmospheric Administration (NOAA) space weather scales (\citet{dpt:12}), a G4 level (K$_p$=8, severe) geomagnetic storm was observed at 23:50 UT on September 7, 2017. There were two more at 01:51 UT and at 13:04 UT on September 8, 2017, as a result of a CME arrival on September 6, 2017. The CME event continued till September 7, 2017. 
Fig.15 shows VTEC (in TECU) plotted as a function of UT (in hours) during the period of September 5-9, 2017, according to latitudes going from north to south i.e Lucknow, Indore, Hyderabad, and Bengaluru along with the values of the monthly mean TEC of the respective stations.
Although, the Dst dropped to a minimum value of -124 nT at 02:00 UT on September 8, 2017, as observed in Fig. 14, the maximum TEC enhancement (Fig.15), among these days, was observed on September 7, 2017, from all the stations. NavIC observed a rise in TEC value in accordance with the GPS observations. The peak TEC on September 7, 2017, from NavIC and GPS, were 67.63 TECU and 66.78 TECU respectively, thus establishing itself to be reliable in monitoring geomagnetic storms. Fig.15 also indicates a TEC enhancement of about 20-33 TECU over the quiet time monthly mean values (plotted in black) at Lucknow, 10-20 TECU enhancement at Indore, 2-10 TECU from Hyderabad and 4-11 TECU from Bengaluru. The monthly mean value of TEC is comparatively low for near equator locations than the off-equatorial stations. The presence of EIA was responsible for the increased TEC at Indore over the near equator locations. The effect of this geomagnetic storm is also more pronounced in the EIA crest region than low latitude locations. One can also note the occurrence of peak TEC is slightly shifted towards later local times as one moves from near equator to off-equatorial sites.

\subsubsection{Storm of September 25-30, 2017}

A G3 level (K$_p$=7, strong) geomagnetic storm, according to NOAA space weather scales, was observed on September 28, 2017. G3 level reached at 05:59 UT on September 28, 2017 and again at 08:05 UT on the same day. Fig.16, similar to Fig.15, shows VTEC plotted as a function of UT during the period of September 25-30, 2017, (It is to be noted that there was a data gap in Lucknow during this period). Dst dropped to a minimum on September 28, 2017, with the value of -55 nT at 07:00 UT as observed in Fig.14 and the peak diurnal maximum of TEC (Fig. 16), among these days, had been observed on the same day from all the stations. On September 28, 2017, TEC values over Indore, from NavIC and GPS were 75.73 and 77.61 TECU respectively, further supporting the reliability of NavIC in monitoring geomagnetic storms. It can also be seen that at Indore, the TEC enhancement is of the order of 10-35 TECU from that of quiet time monthly mean values, 16-18 TECU over Hyderabad and 2-25 TECU over Bengaluru during this storm. From both Fig.15 and 16, it can be observed that NavIC observables were able to capture the geomagnetic storms similar to GPS, thereby proving itself to be reliable in monitoring space weather events.

\section{Discussions and Conclusion}

NavIC is one of the recent regional navigation satellite system launched specifically for the Indian subcontinent. Because of the availability of three geostationary satellites, the part of the ionosphere can be monitored continuously. The additional ray path through the ionosphere due to this system can also help to monitor space weather effects more effectively when used with other GNSS constellations. it is observed that one of the NavIC satellite (PRN-5) measures a part of the ionosphere where the none of GPS ray paths are available in surrounding 1\ensuremath{^{\circ}} x 1\ensuremath{^{\circ}} area. This highlights another advantage of assimilating NavIC data with GPS data for ionospheric studies over the Indian subcontinent because of the continuous availability of NavIC signals from all satellites throughout the day.

This paper further reveals the performance of the NavIC for studying the space weather, evaluated at Indore using one year's observations. To study the performance of NavIC during severe space weather events, detailed case studies of two geomagnetic storms are also presented. It has been observed that the diurnal variation of TEC recorded from the NavIC matches closely with that of GPS. However, the observation of a systematic anomaly in NavIC data and may be due to the instrumental effect which can be further investigated for a better understanding. The corrected results indicate a consistency between GPS and NavIC which establishes NavIC as a reliable system to probe the ionosphere. However, the analysis presented in the paper is limited to one station of NavIC i.e., Indore, but similar comparative analyses with NavIC at various stations can further be implemented in order to validate the performance of this constellation with respect to the Indian region. 
To study the performance of the NavIC system in response to extreme space weather events, data w.r.t two geomagnetic storms were analyzed. The enhancement of TEC was observed in both GPS and NavIC signals from different locations over India during the moderate storm on September 28, 2017, and the intense storm on September 8, 2017, which further confirms that NavIC receivers are reliable in monitoring ionospheric response to geomagnetic storms. In both the cases, TEC estimated by NavIC matches with that of GPS. The study confirms the reliability of the NavIC data for ionospheric studies and will be helpful to utilize NavIC for such studies over the equatorial ionosphere across the Indian subcontinent.

\section*{Acknowledgments} 

DA acknowledges the Department of Science and Technology for providing her with the INSPIRE fellowship grant to pursue her research. SC acknowledges Space Applications Centre (SAC), ISRO for providing research fellowship under NavIC/GAGAN utilization program: NGP-17. SAC, ISRO is further acknowledged by the authors for providing the NavIC receiver (ACCORD) under NGP-17 to the Discipline of Astronomy, Astrophysics and Space Engineering, IIT Indore. SD acknowledges the financial support received under the INSPIRE Faculty Scheme (DST/INSPIRE/04/2014/002492) and ISRO NGP-1. The authors would also like to acknowledge Prof. Gopi Seemala of the Indian Institute of Geomagnetism (IIG), Navi Mumbai, India for providing the software to analyze the IGS data.

\newpage\bibliographystyle{model2-names.bst}\biboptions{authoryear}
\bibliography{sample.bib}

\newpage
\begin{figure}
\centering
\includegraphics[width=6in,height=6in]{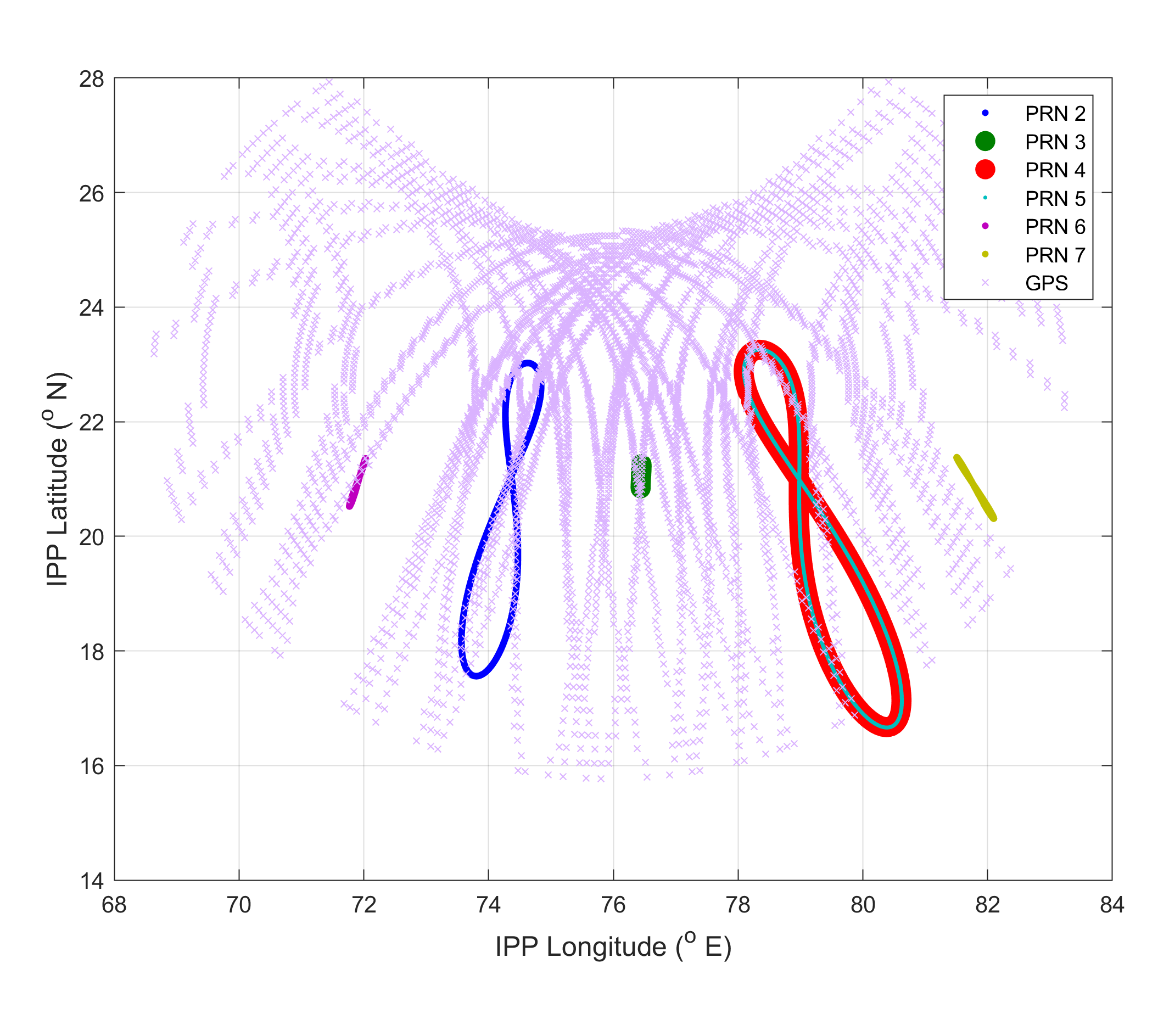}
\caption{Latitude vs Longitude plot of  Ionospheric Pierce Points (IPP) for all NavIC and GPS satellites observed at DAASE, IIT Indore.}
\end{figure}

\begin{figure}
\centering
\includegraphics[width=6in,height=4in]{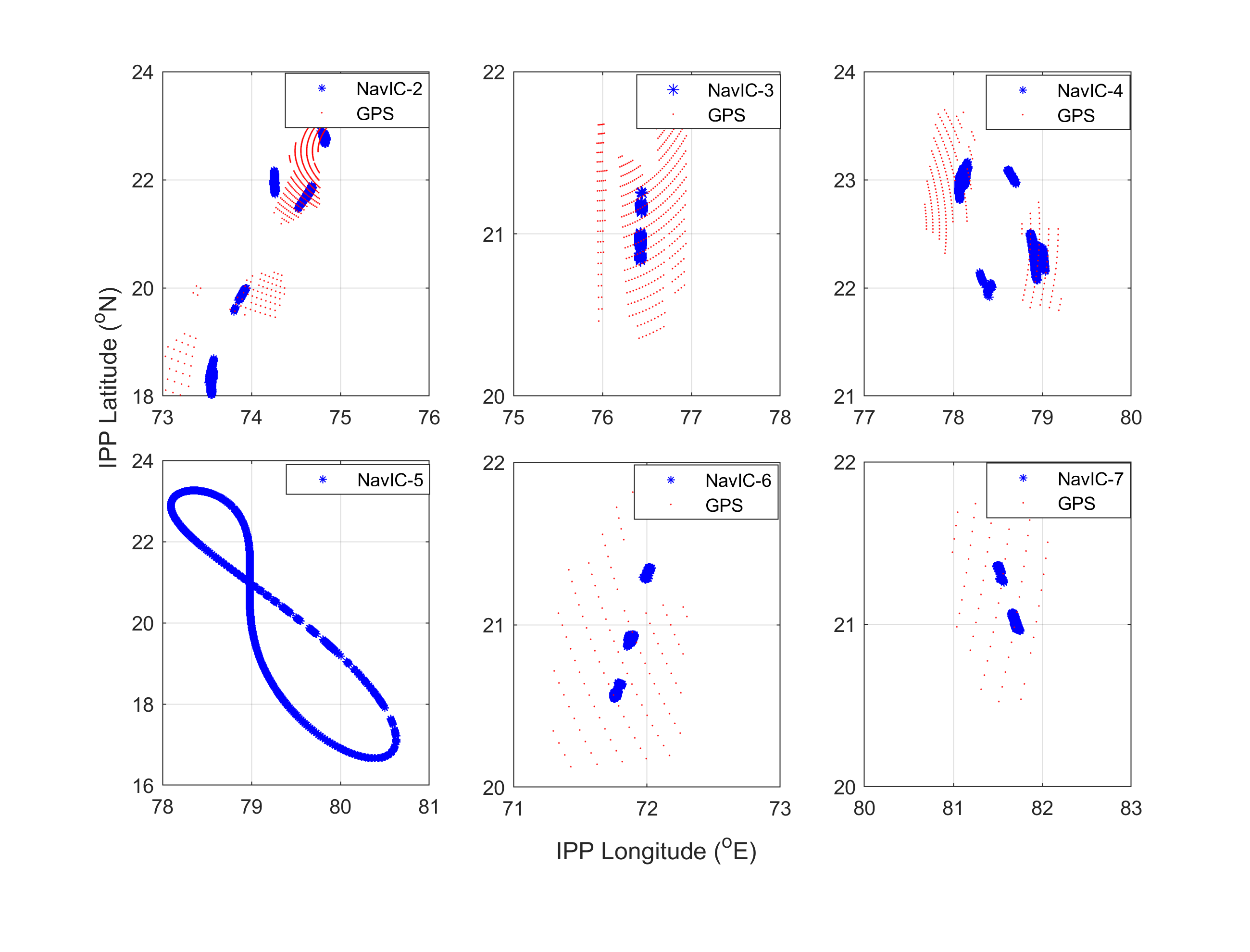}
\caption{The GPS (red) ray paths in the  $1^{\circ}$x$1^{\circ}$ grid surrounding the IPP for each of the NavIC satellite (NavIC-2 to 7(blue))}
\end{figure}

\newpage
\begin{figure} 
\centering
    \includegraphics[width=3.5in,height=2in]{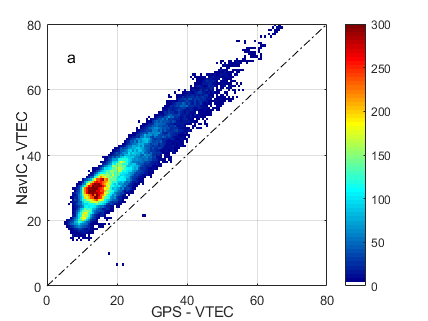}
    \includegraphics[width=3.5in,height=2in]{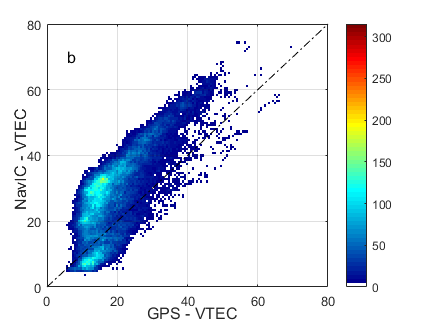}
    \includegraphics[width=3.5in,height=2in]{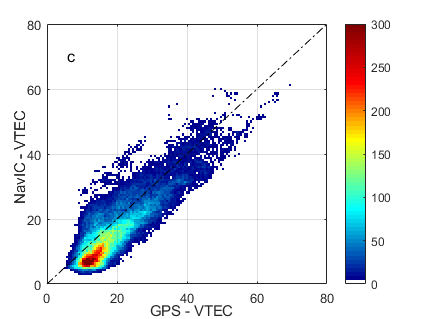}\\
    \caption{The scatter plot between NavIC and GPS derived VTEC for a period of one year (a) VTEC estimates without any corrections (b) VTEC estimates after making diurnal minimum value corrections without removal of anomalous data (c) VTEC estimates after making diurnal minimum value corrections with removal of anomalous data. Color bar indicates the number of data points.}
\end{figure}

\newpage
\begin{figure}
\centering 
\includegraphics[width=3.5in,height=2in]{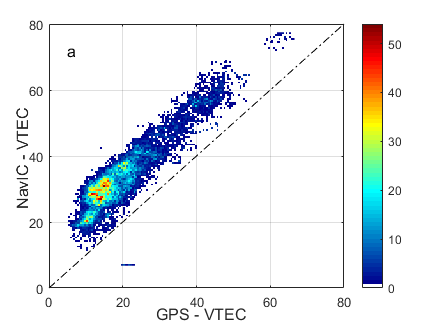}
\includegraphics[width=3.5in,height=2in]{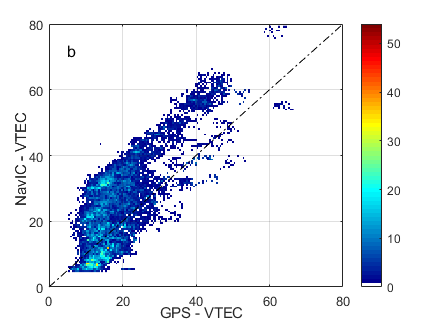}
\includegraphics[width=3.5in,height=2in]{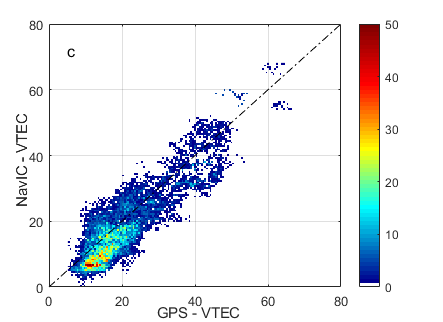}\\
    \caption{The scatter plot between NavIC and GPS derived VTEC for the quiet days.(a) VTEC estimates without any corrections (b) VTEC estimates after making diurnal minimum value corrections without removal of anomalous data (c) VTEC estimates after making diurnal minimum value corrections with removal anomalous data. Color bar indicates the number of data points.}
\end{figure}

\newpage
\begin{figure}
\centering
\includegraphics[width=3.5in,height=2in]{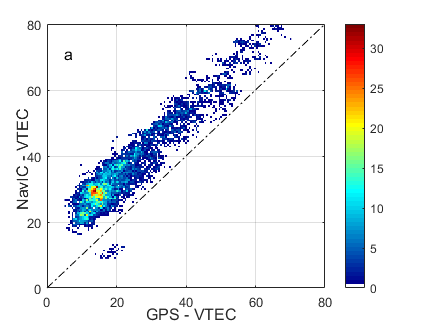}
\includegraphics[width=3.5in,height=2in]{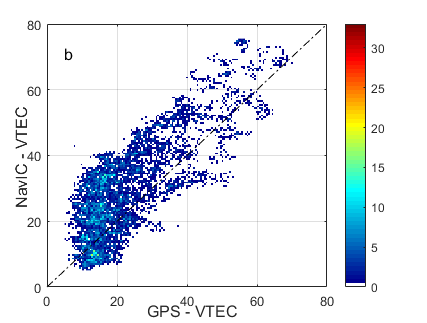}
\includegraphics[width=3.5in,height=2in]{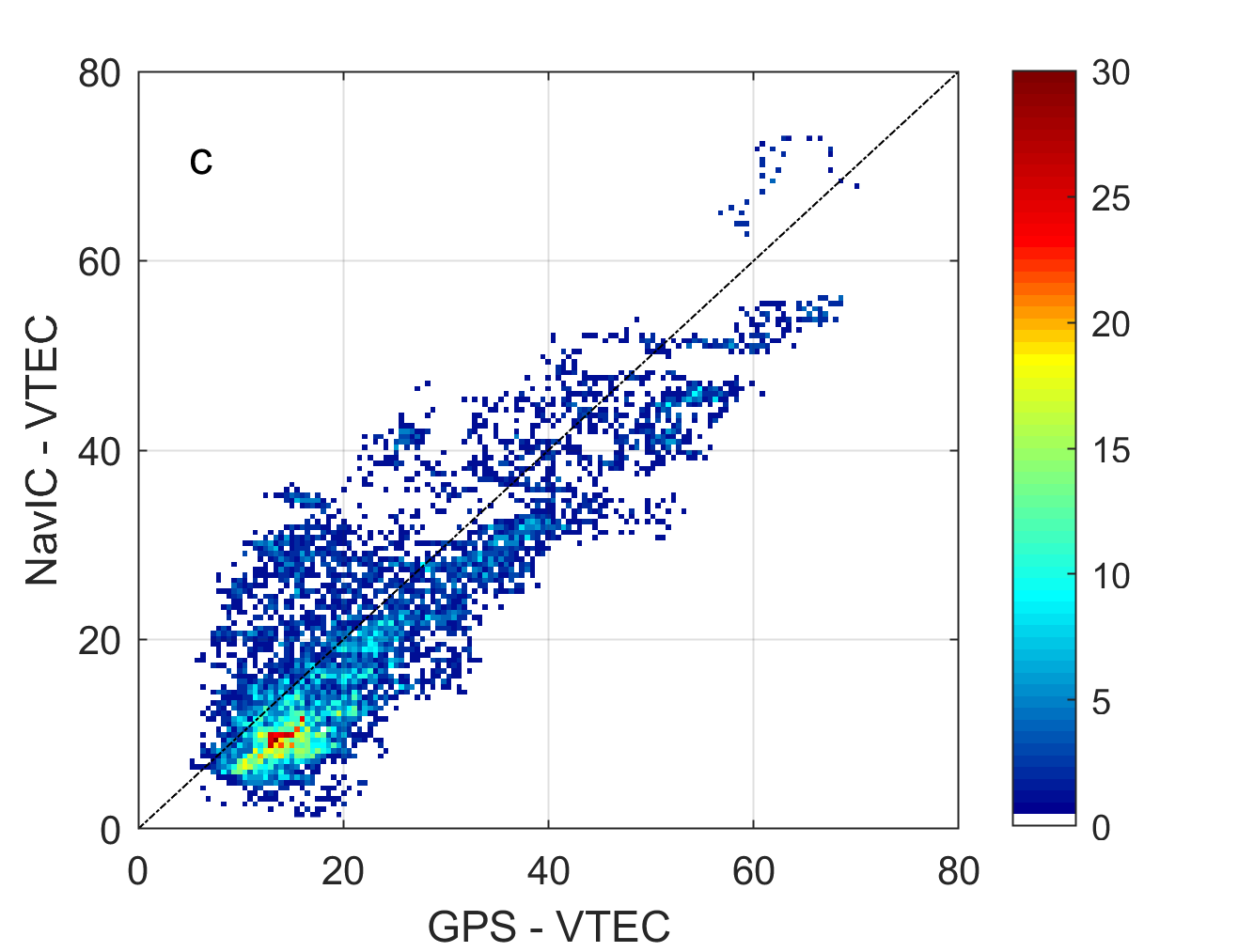}\\
    \caption{The scatter plot between NavIC and GPS derived VTEC for the disturbed days.(a) VTEC estimates without any corrections (b) VTEC estimates after making diurnal minimum value corrections without removal of anomalous data (c) VTEC estimates after making diurnal minimum value corrections with removal anomalous data. Color bar indicates the number of data points.}
\end{figure}

\begin{figure}
\centering
\includegraphics[width=5in,height=3in]{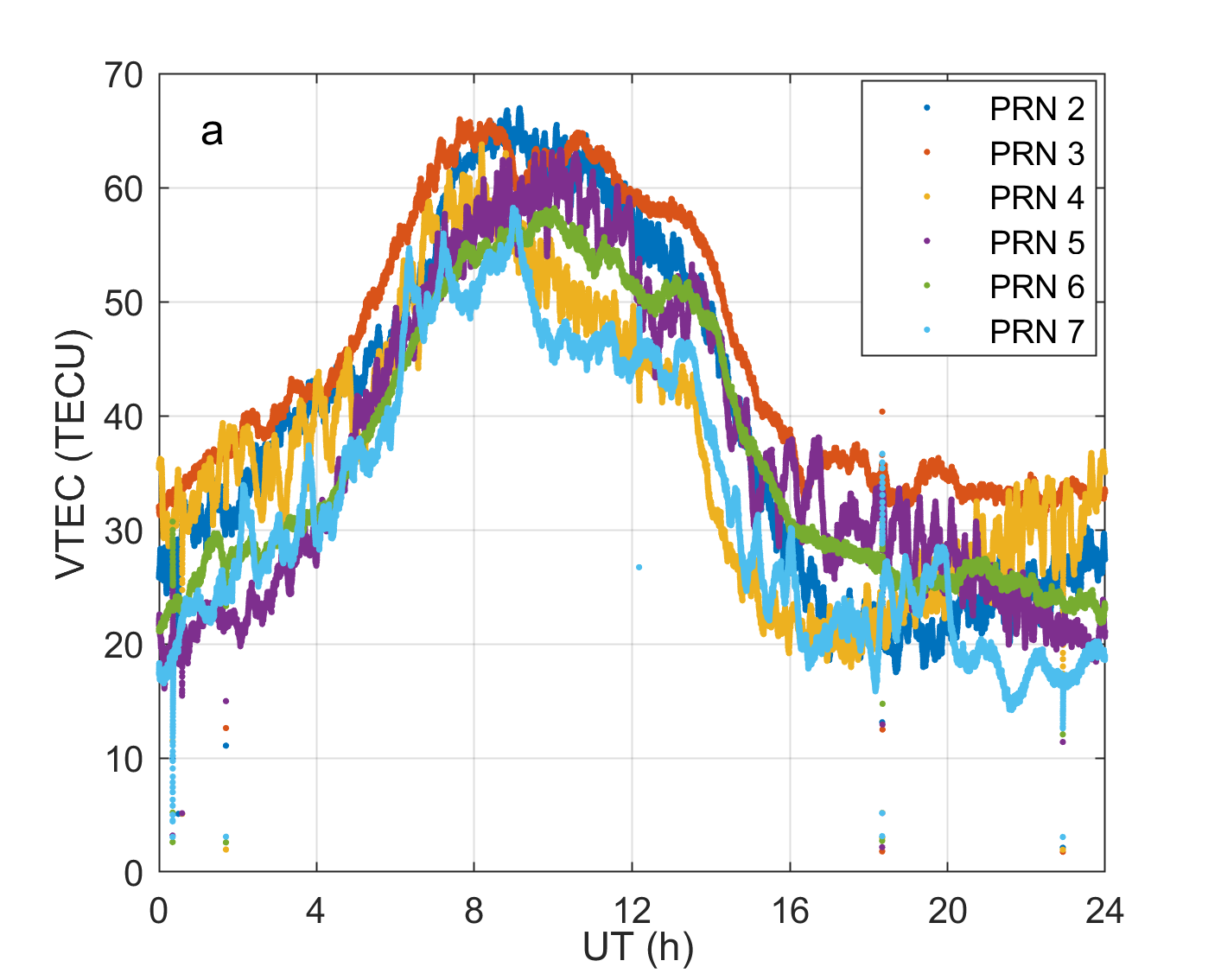}
\includegraphics[width=5in,height=3in]{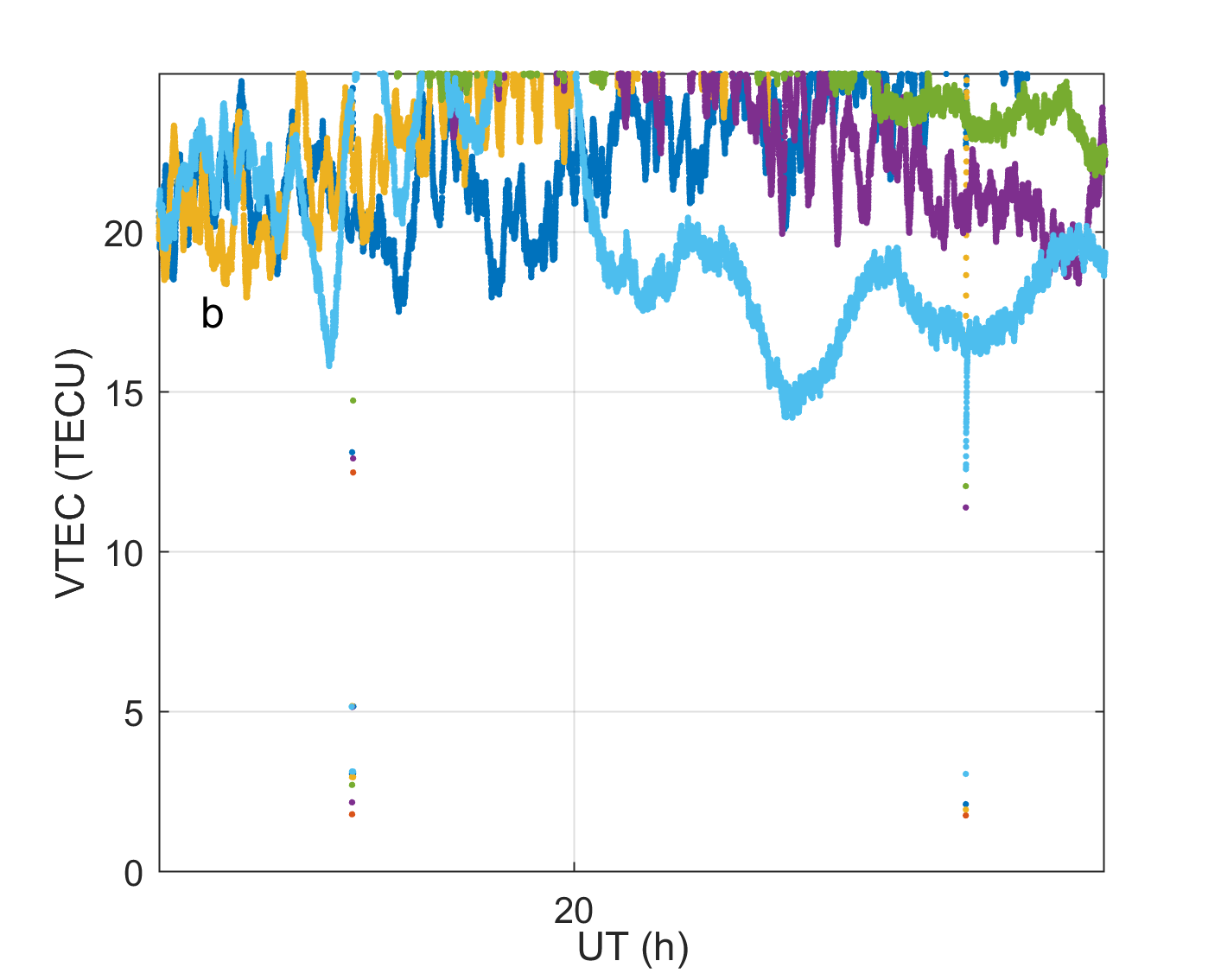}\\
    \caption{The anomalous data spotted in NavIC VTEC. (a) The diurnal pattern of VTEC of NavIC (b) The zoomed-in version of (a) representing the anomalous data during quiet ionospheric time (20-24UT)}
\end{figure}

\newpage
\begin{figure}
\centering
\includegraphics[width=2.5in,height=6.5in]{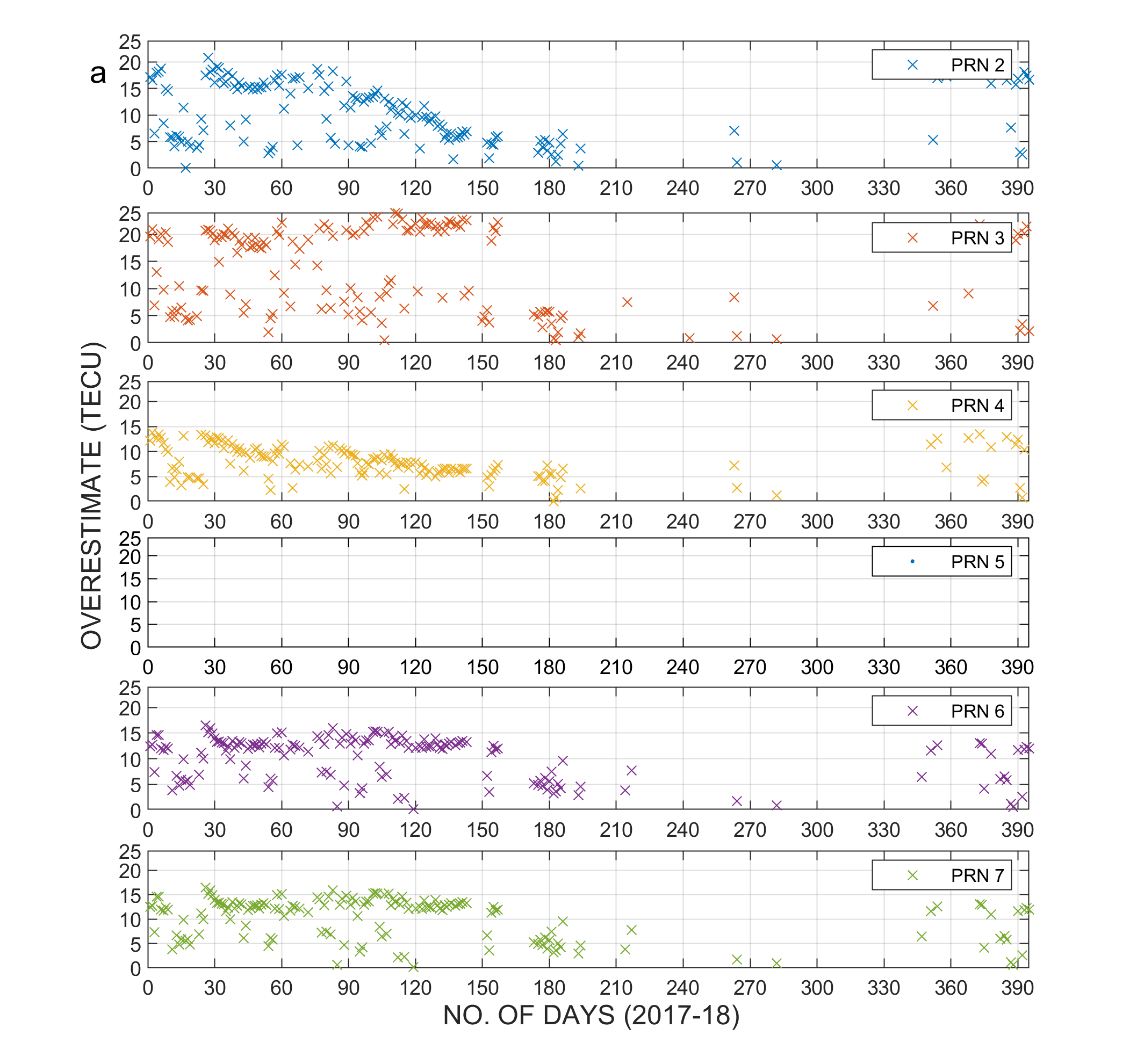}
\includegraphics[width=2.5in,height=6.5in]{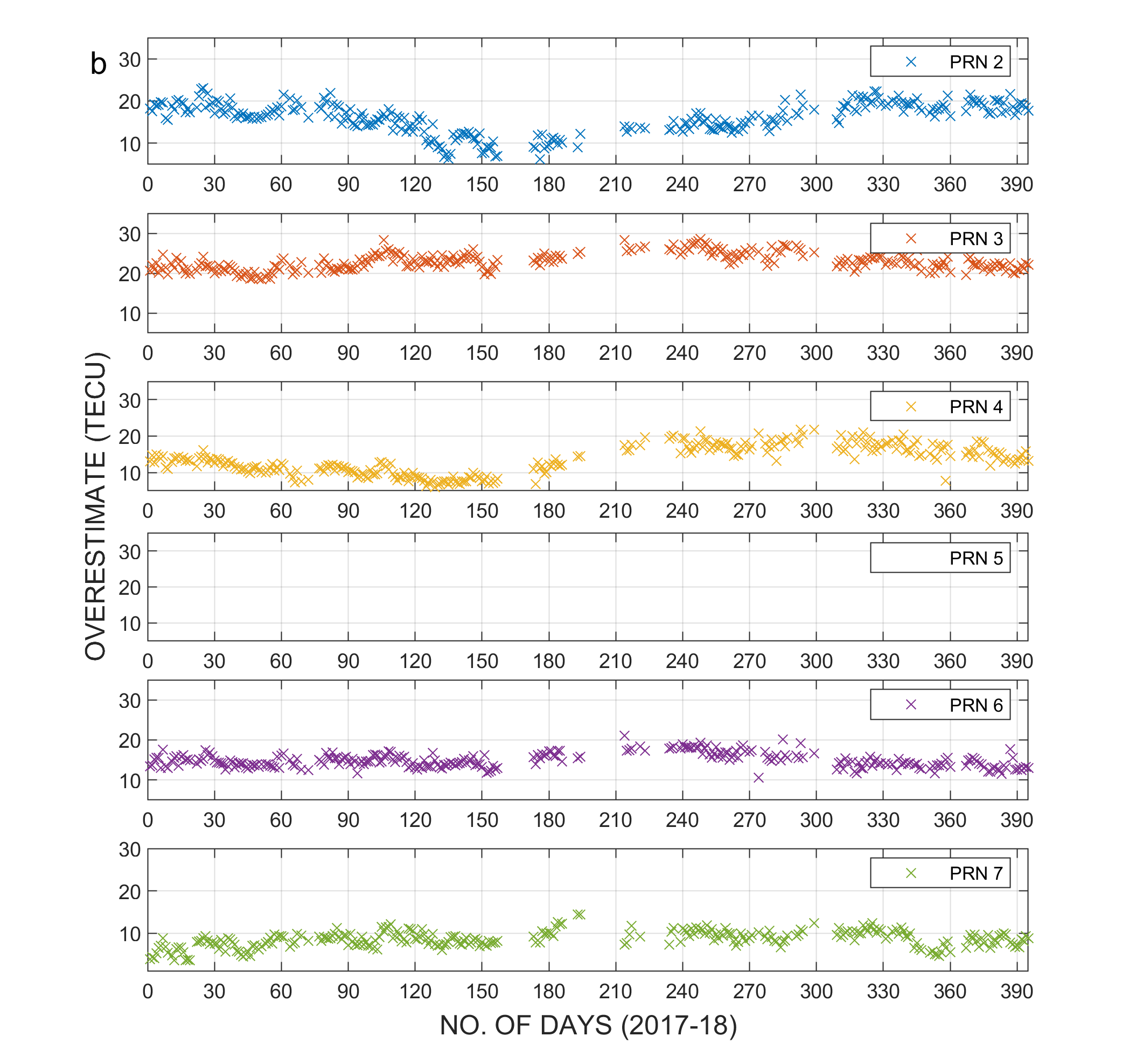}\\
\caption{The magnitude of overestimates in NavIC derived TEC over GPS derived TEC (a) before removal of anomalous data; (b) after removal of anomalous data for individual NavIC satellites.}
\end{figure}

\newpage
\begin{figure}
\centering
\includegraphics[width=1.75in,height=4.25in]{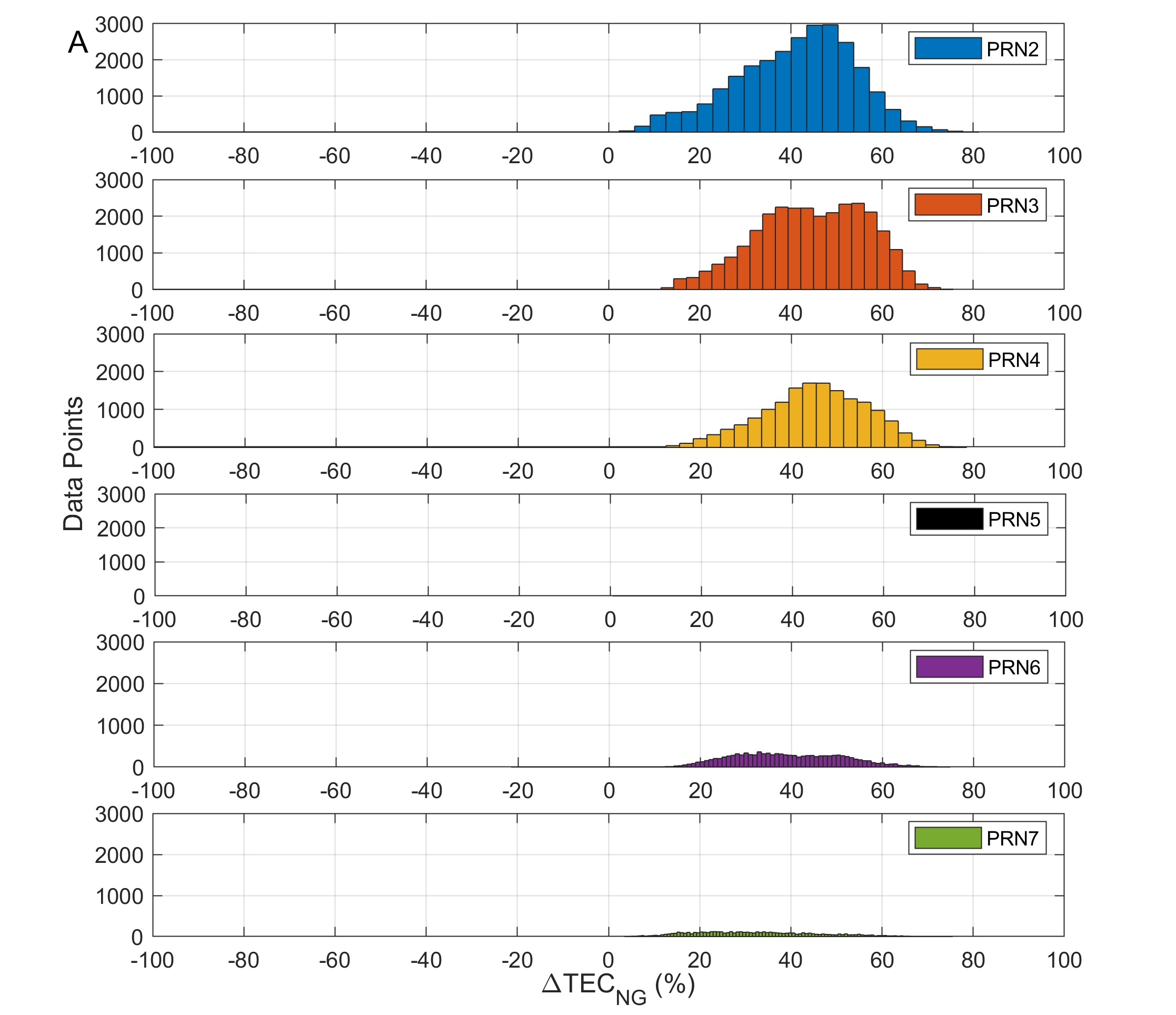}
\includegraphics[width=1.75in,height=4.25in]{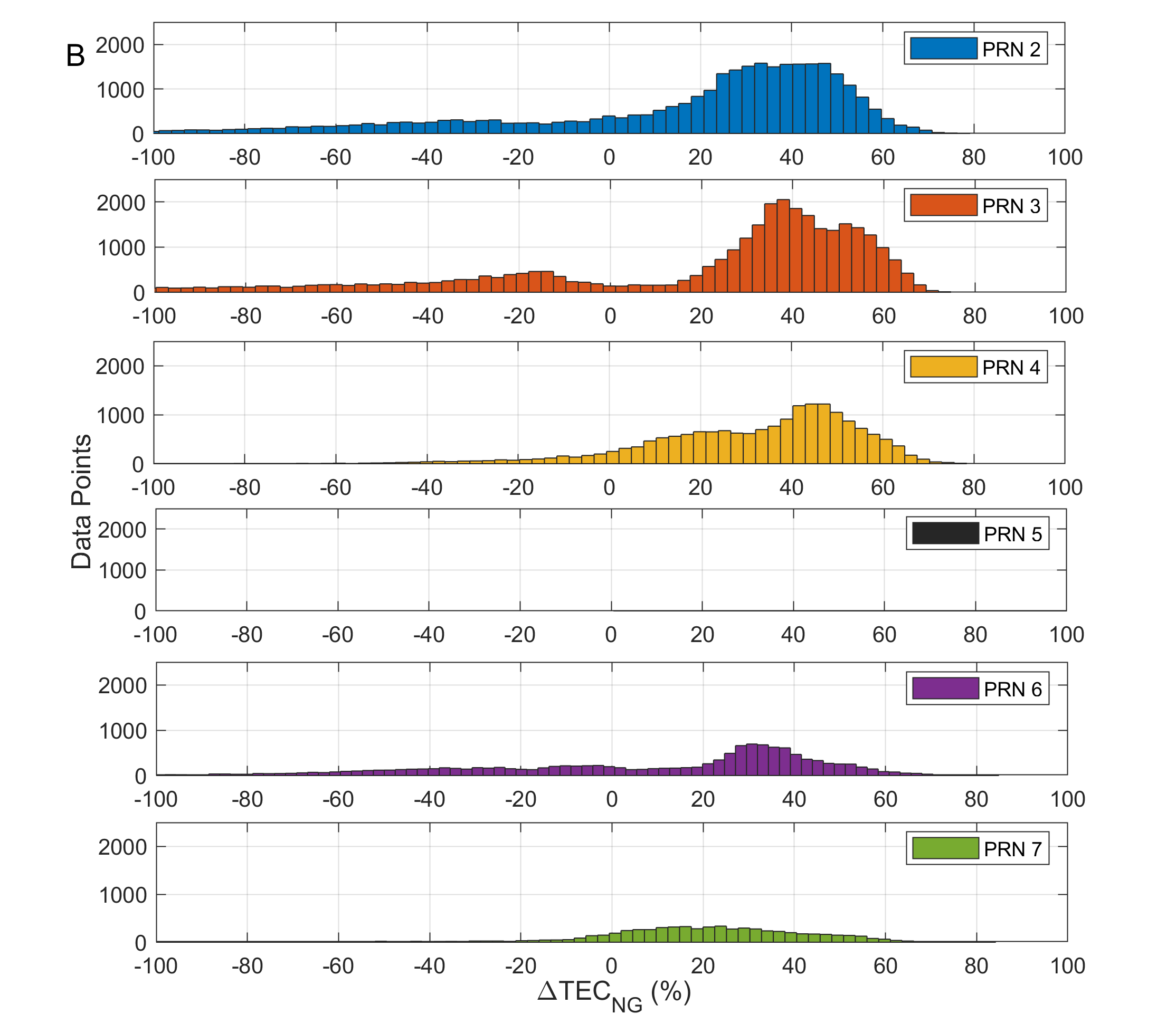}
\includegraphics[width=1.75in,height=4.25in]{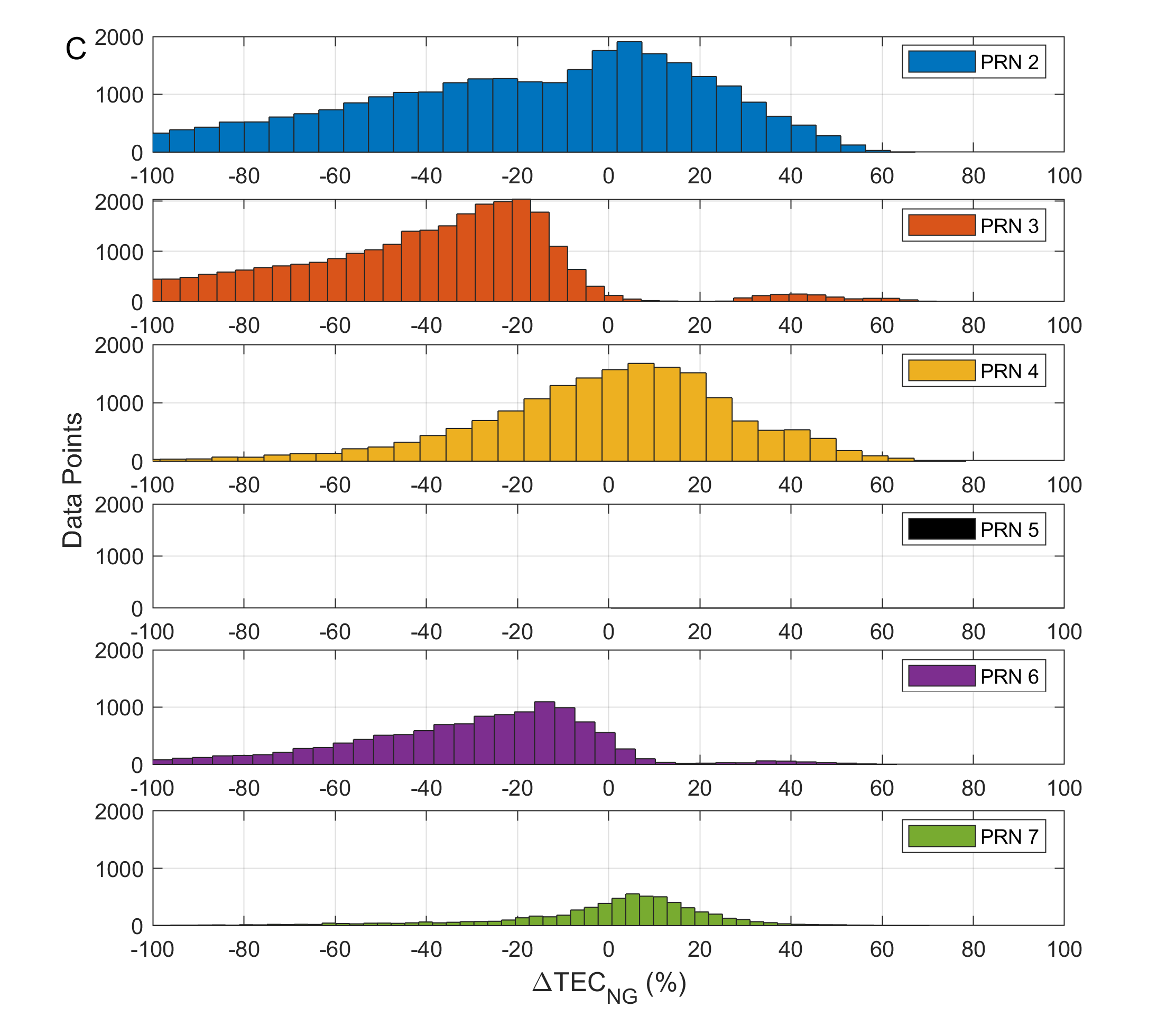}\\
\caption{Distribution ($\Delta TEC_{NG}$(\%)) of all days for each of the NavIC satellites (a) without any corrections (b) after making diurnal minimum value corrections without removal of anomalous data (c) after making diurnal minimum value corrections with anomalous data removed.}
\end{figure}

\newpage
\begin{figure}
\centering
\includegraphics[width=1.75in,height=4.25in]{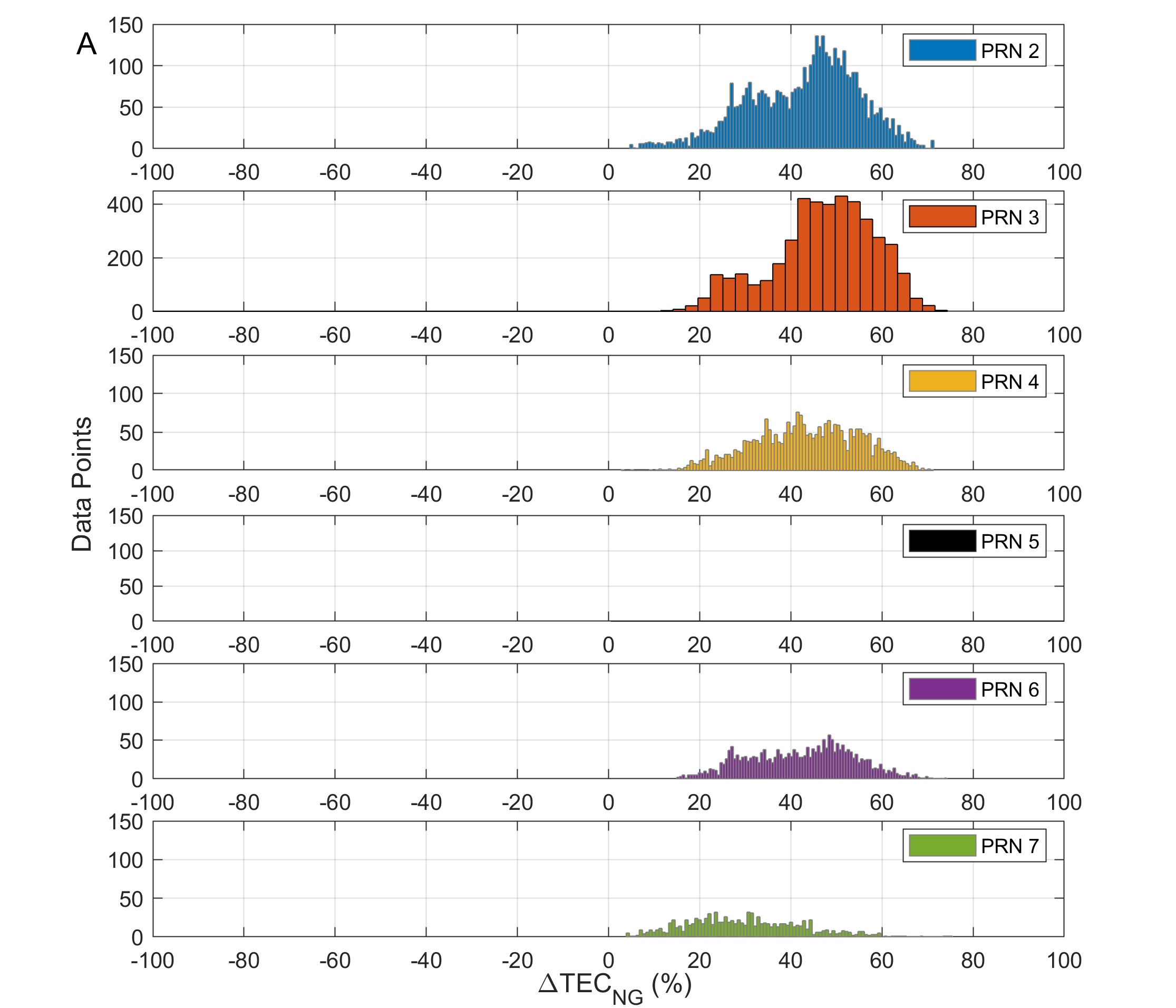}
\includegraphics[width=1.75in,height=4.25in]{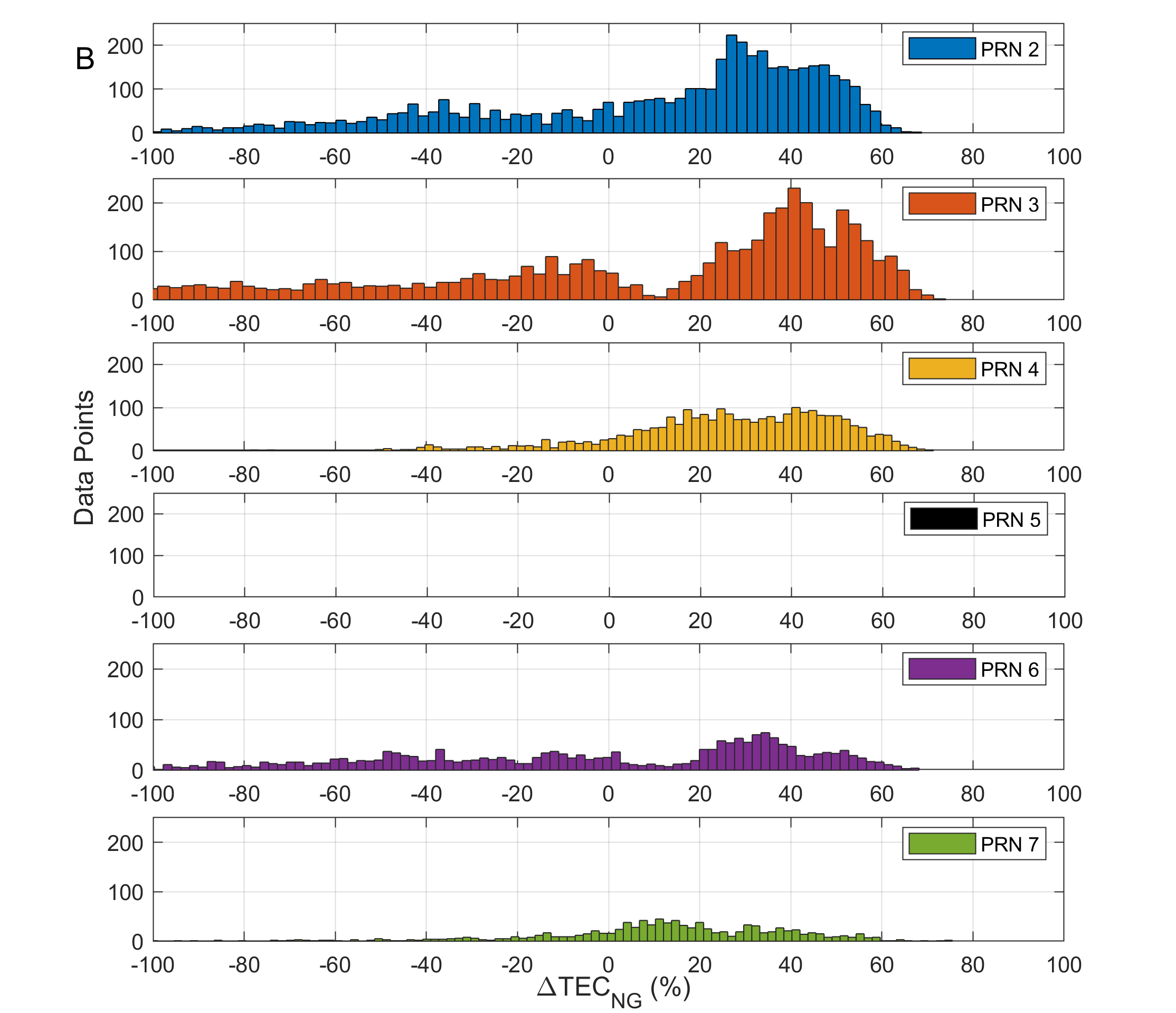}
\includegraphics[width=1.75in,height=4.25in]{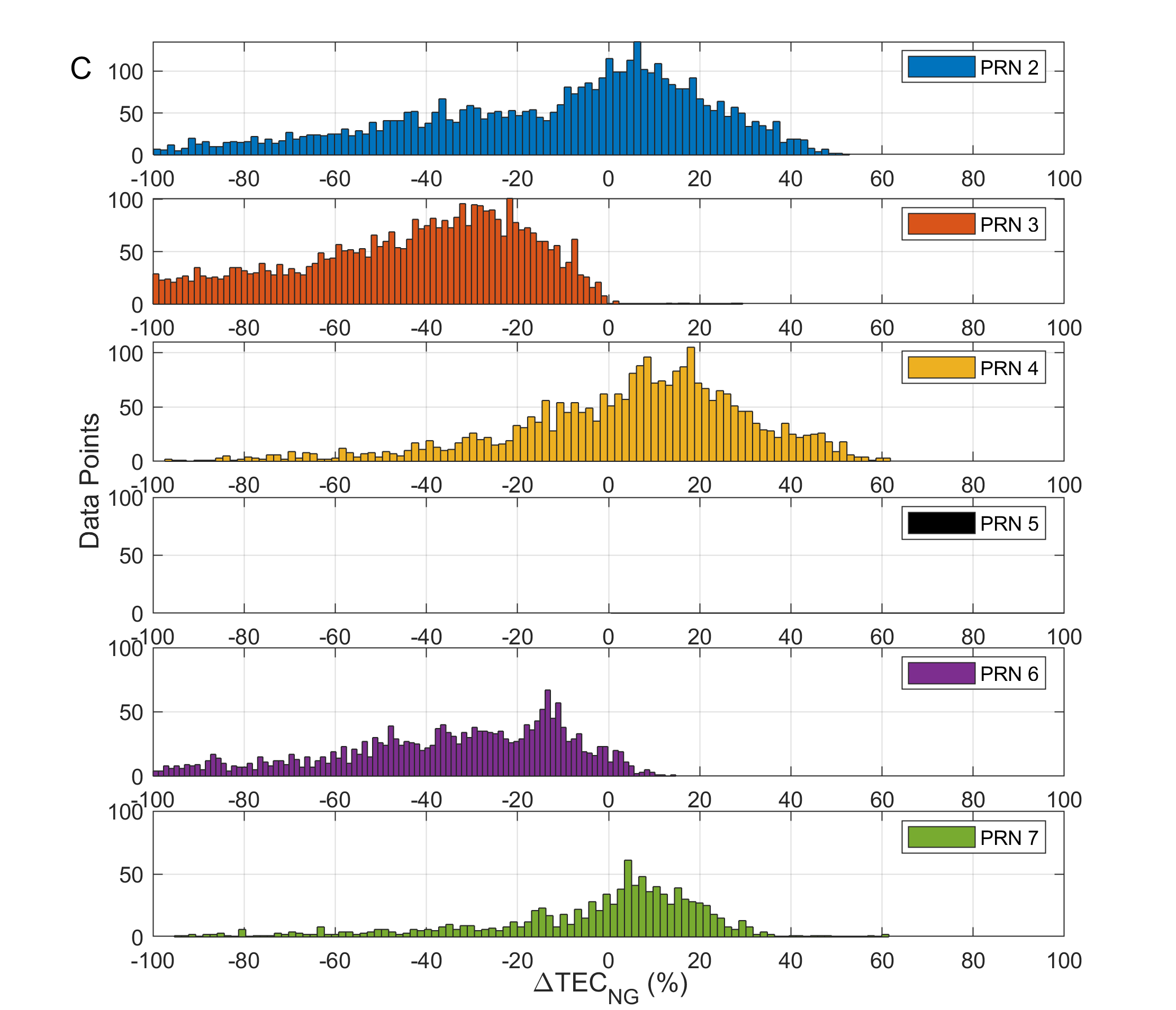}\\
\caption{Distribution ($\Delta TEC_{NG}$(\%)) of Quiet Days for each of the NavIC satellites.(a) without any corrections (b) after making diurnal minimum value corrections without removal of anomalous data (c) after making diurnal minimum value corrections with anomalous data removed.}
\end{figure}

\newpage
\begin{figure}
\centering
\includegraphics[width=1.75in,height=4.25in]{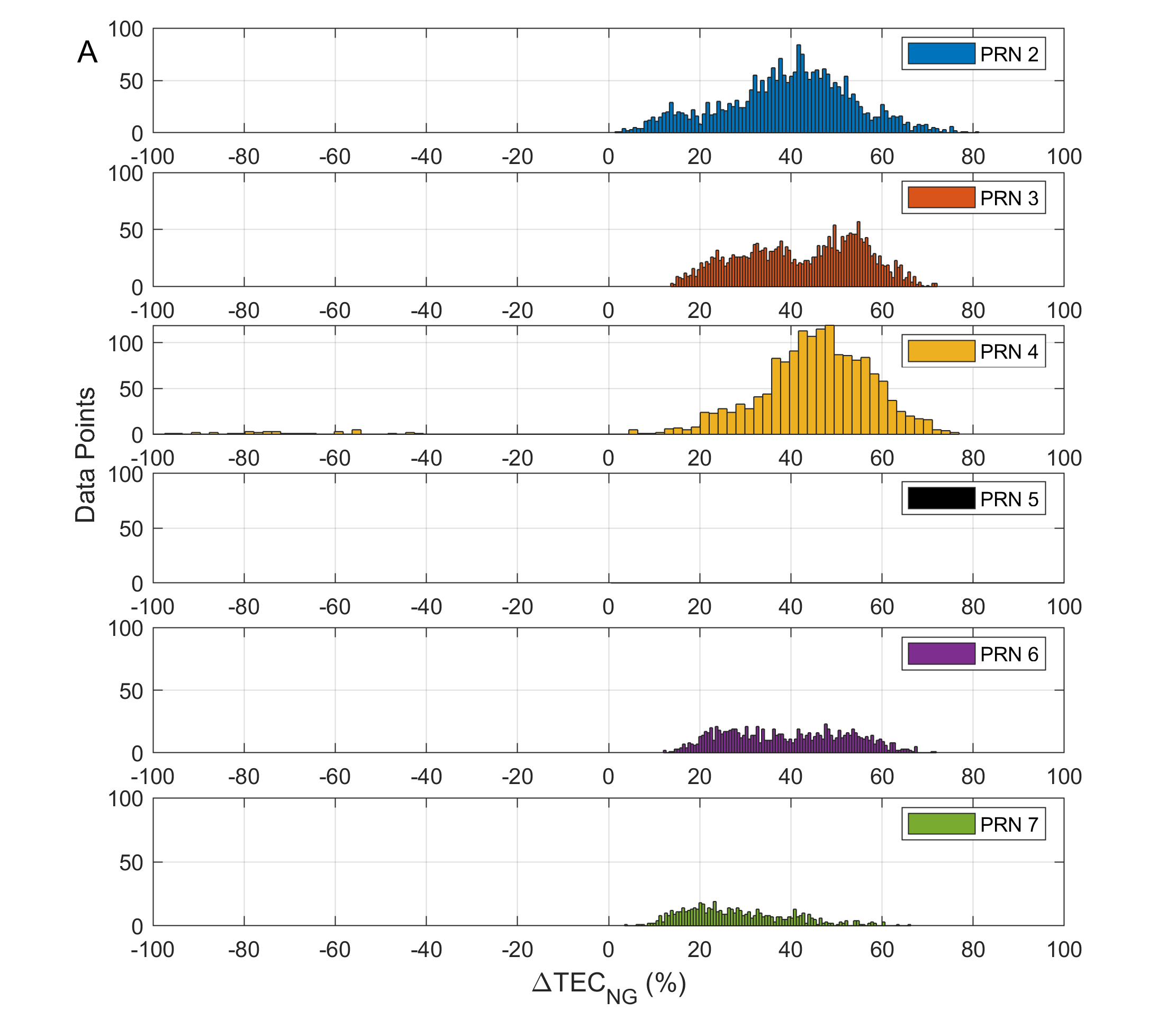}
\includegraphics[width=1.75in,height=4.25in]{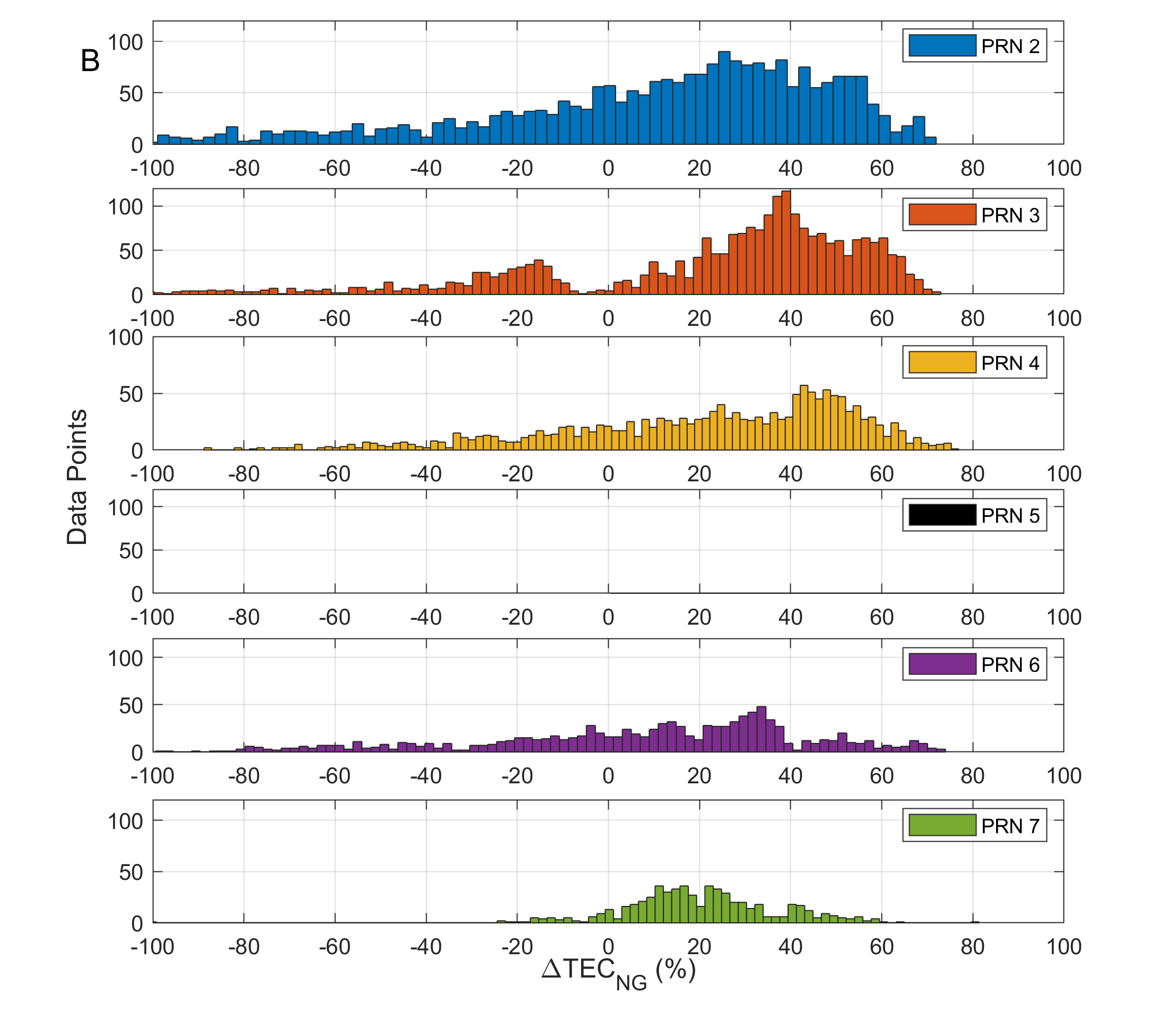}
\includegraphics[width=1.75in,height=4.25in]{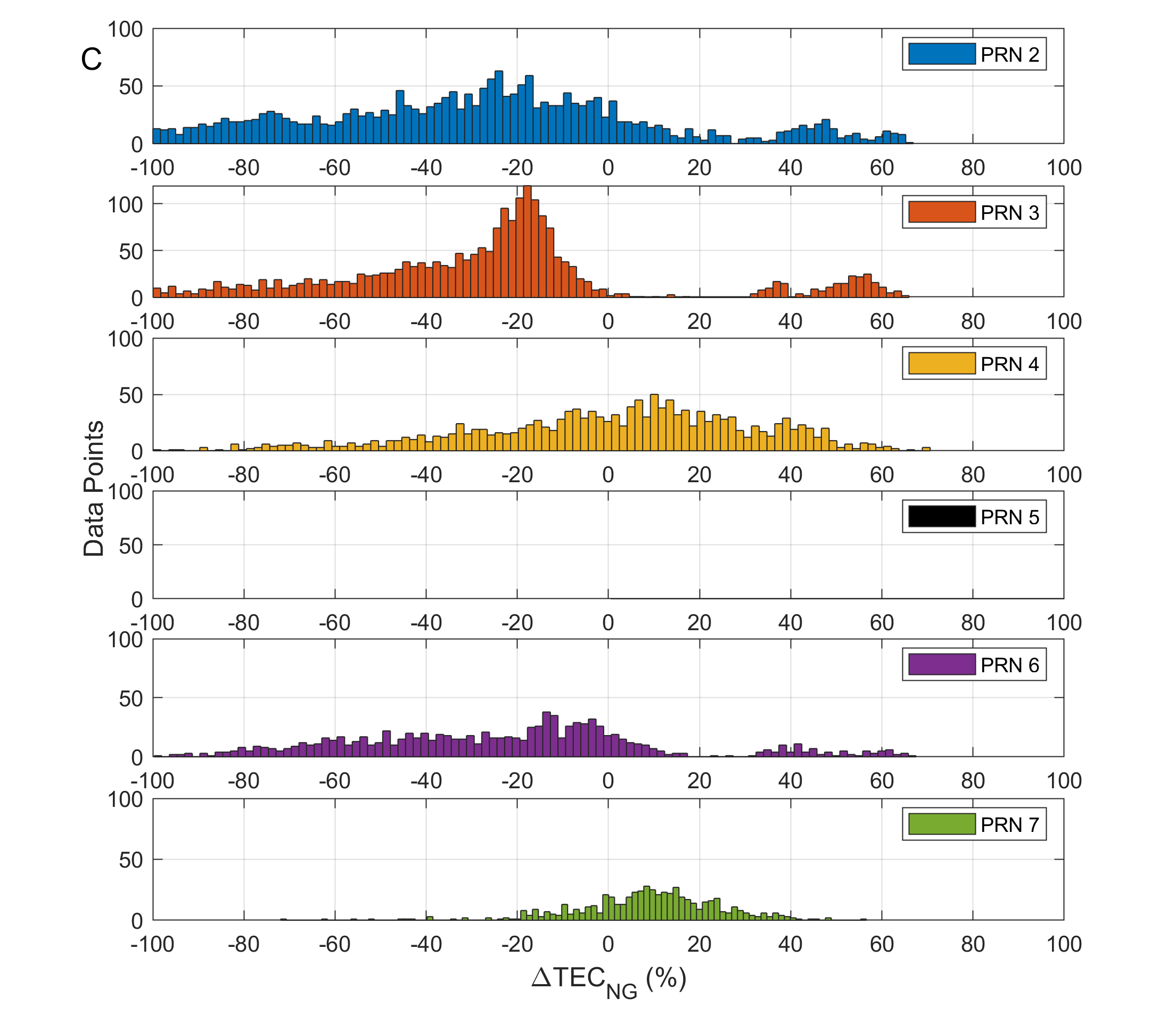}\\
\caption{Distribution ($\Delta TEC_{NG}$(\%)) of Disturbed Days for each of the NavIC satellites.(a) without any corrections (b) after making diurnal minimum value corrections without removal of anomalous data (c) after making diurnal minimum value corrections with anomalous data removed.}
\end{figure}

\newpage
\begin{figure}
\centering
\includegraphics[width=2in,height=2in]{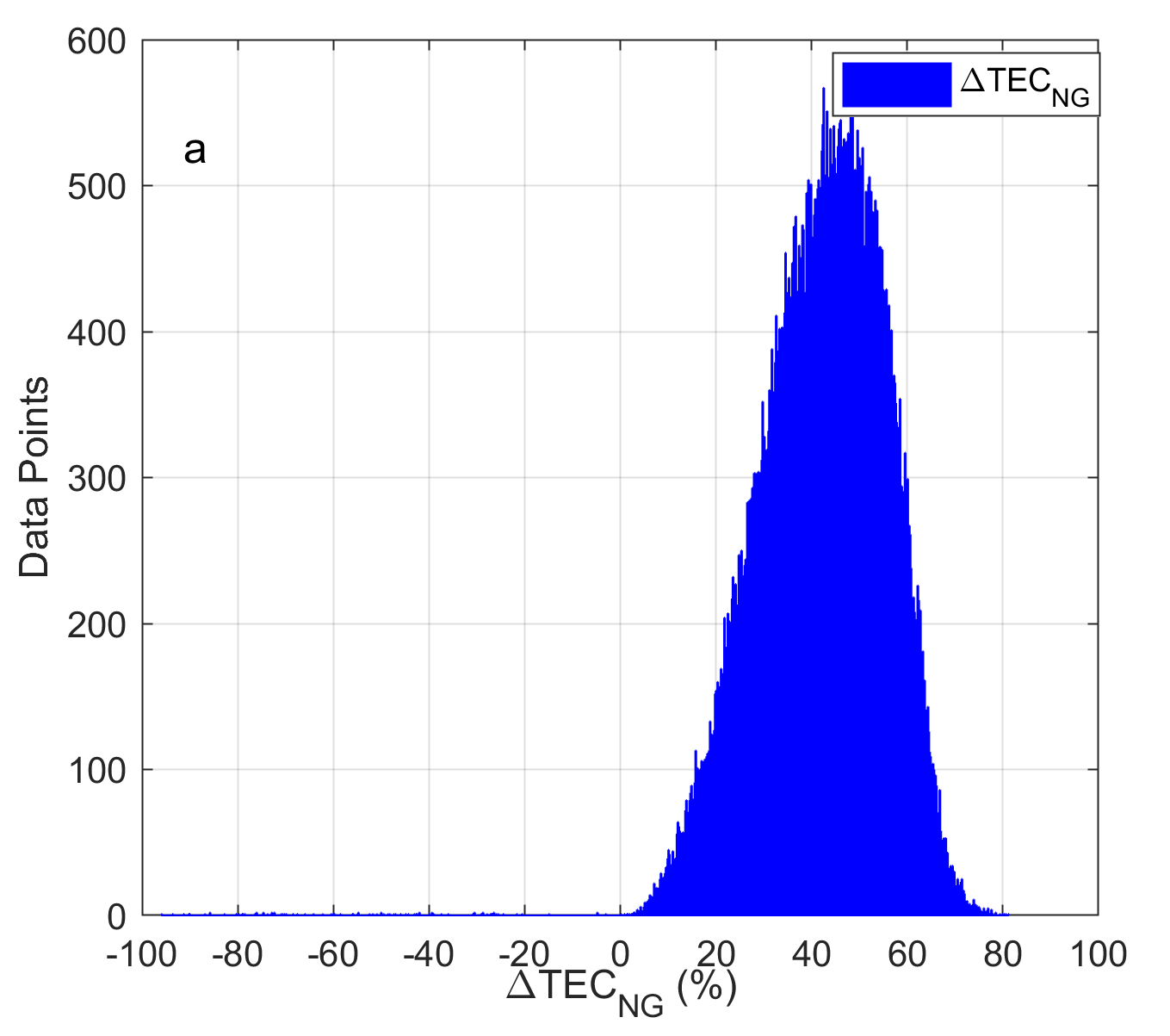}
\includegraphics[width=2in,height=2in]{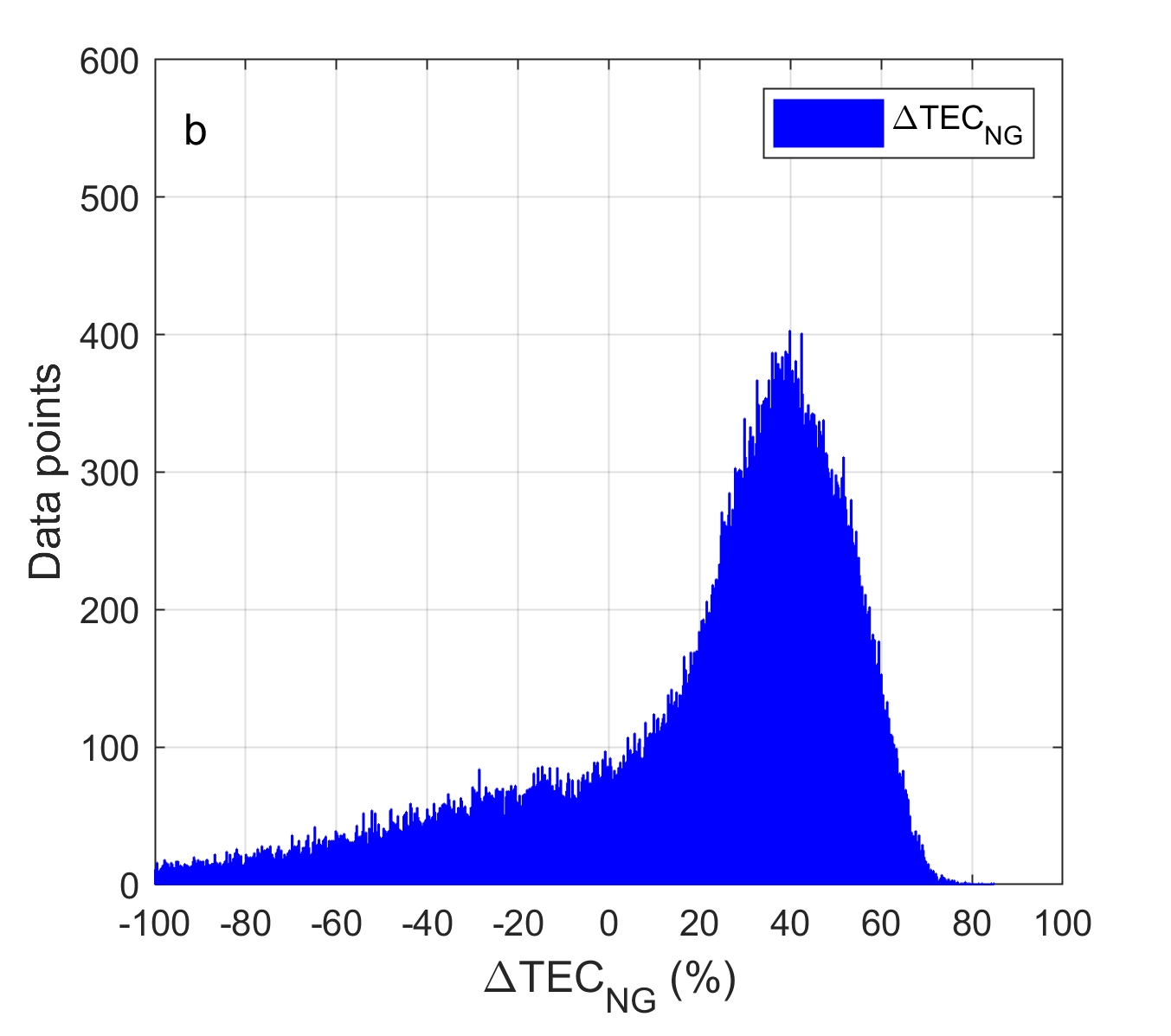}
\includegraphics[width=2in,height=2in]{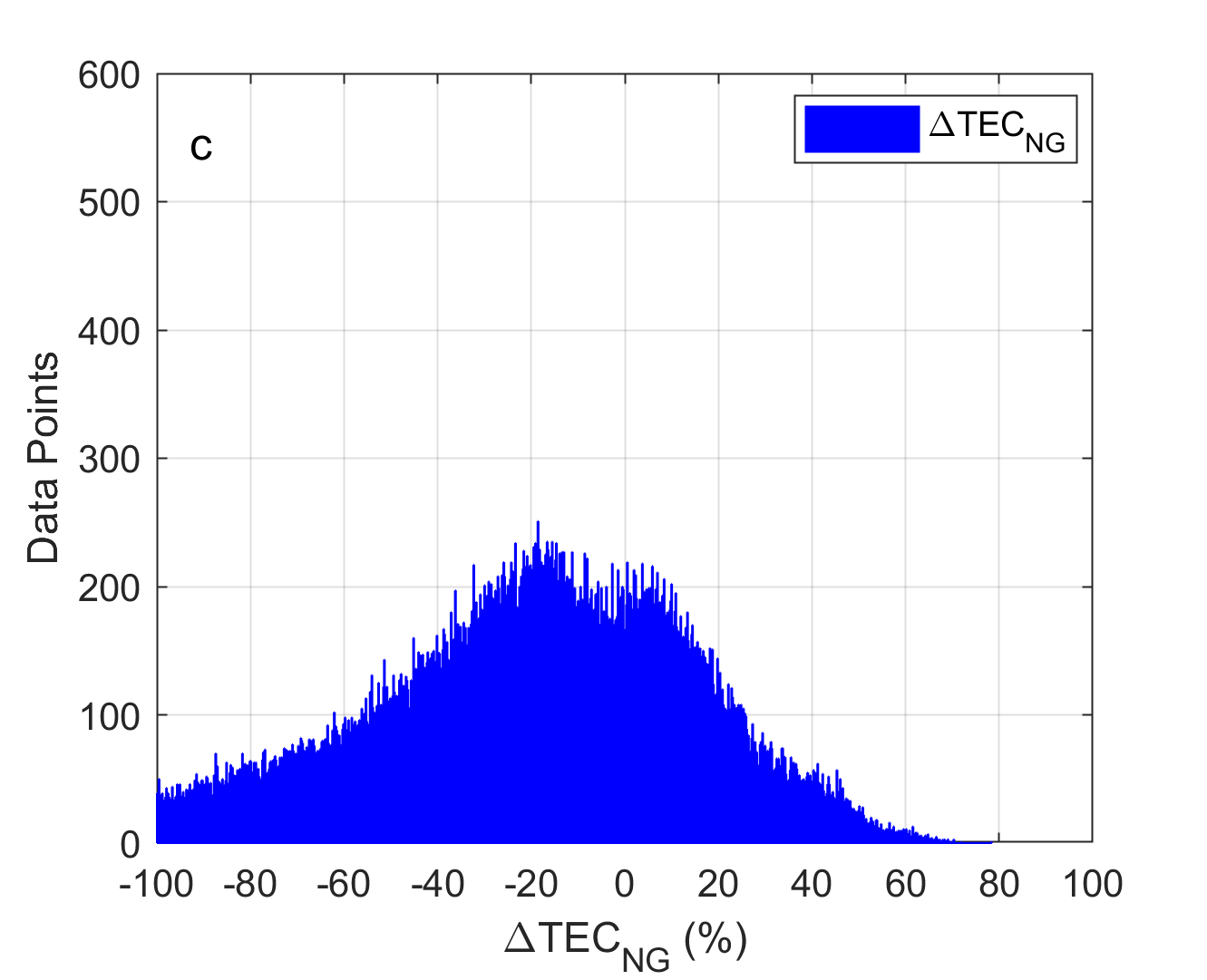}\\
\caption{$\Delta TEC_{NG}$(\%) distribution over the study period combining all NavIC satellites. (a) without any corrections (b) after making diurnal minimum value corrections without removal of anomalous data (c) after making diurnal minimum value corrections with removal of anomalous data}
\end{figure}

\newpage
\begin{figure}
\centering
\includegraphics[width=2in,height=2in]{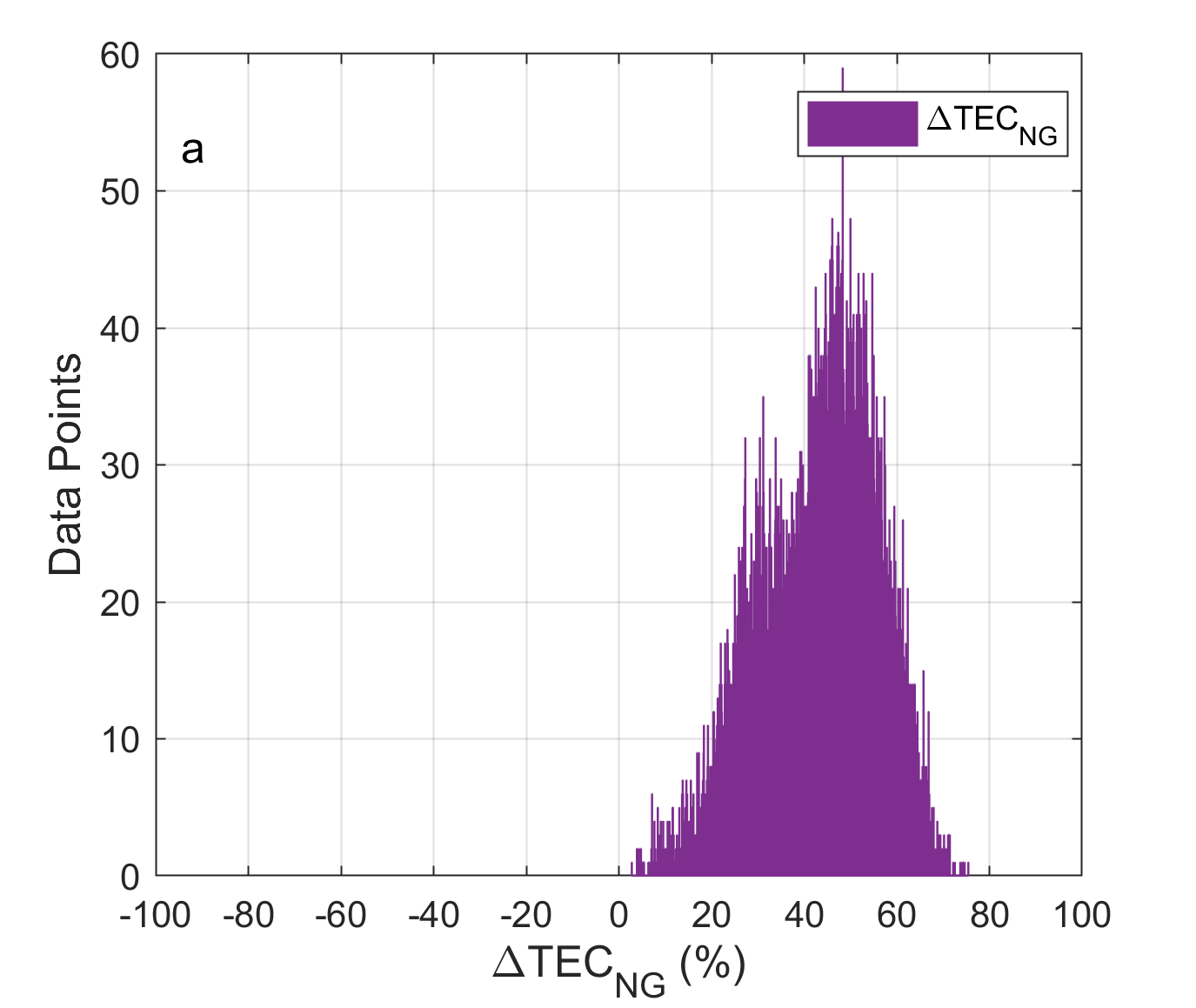}
\includegraphics[width=2in,height=2in]{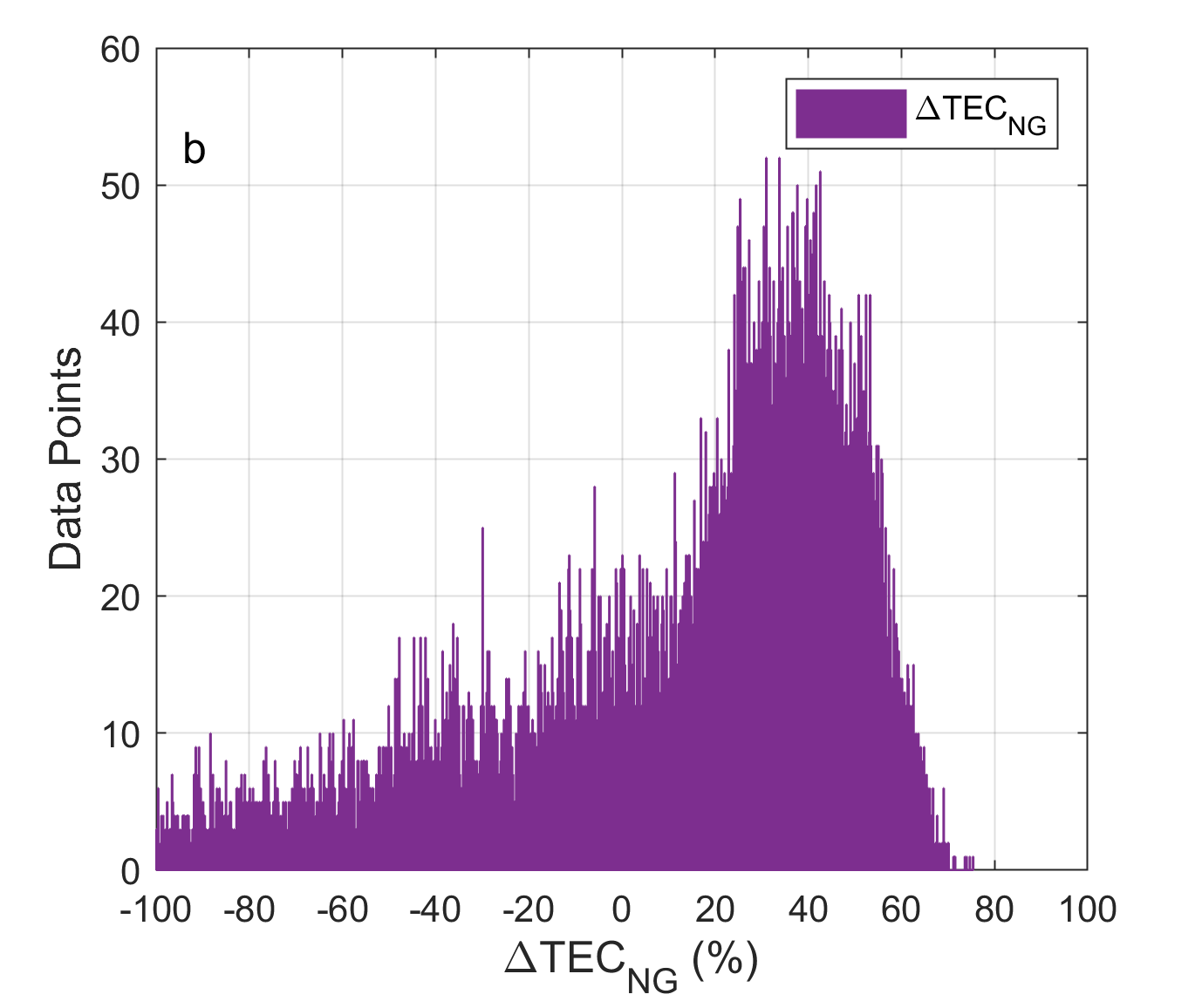}
\includegraphics[width=2in,height=2in]{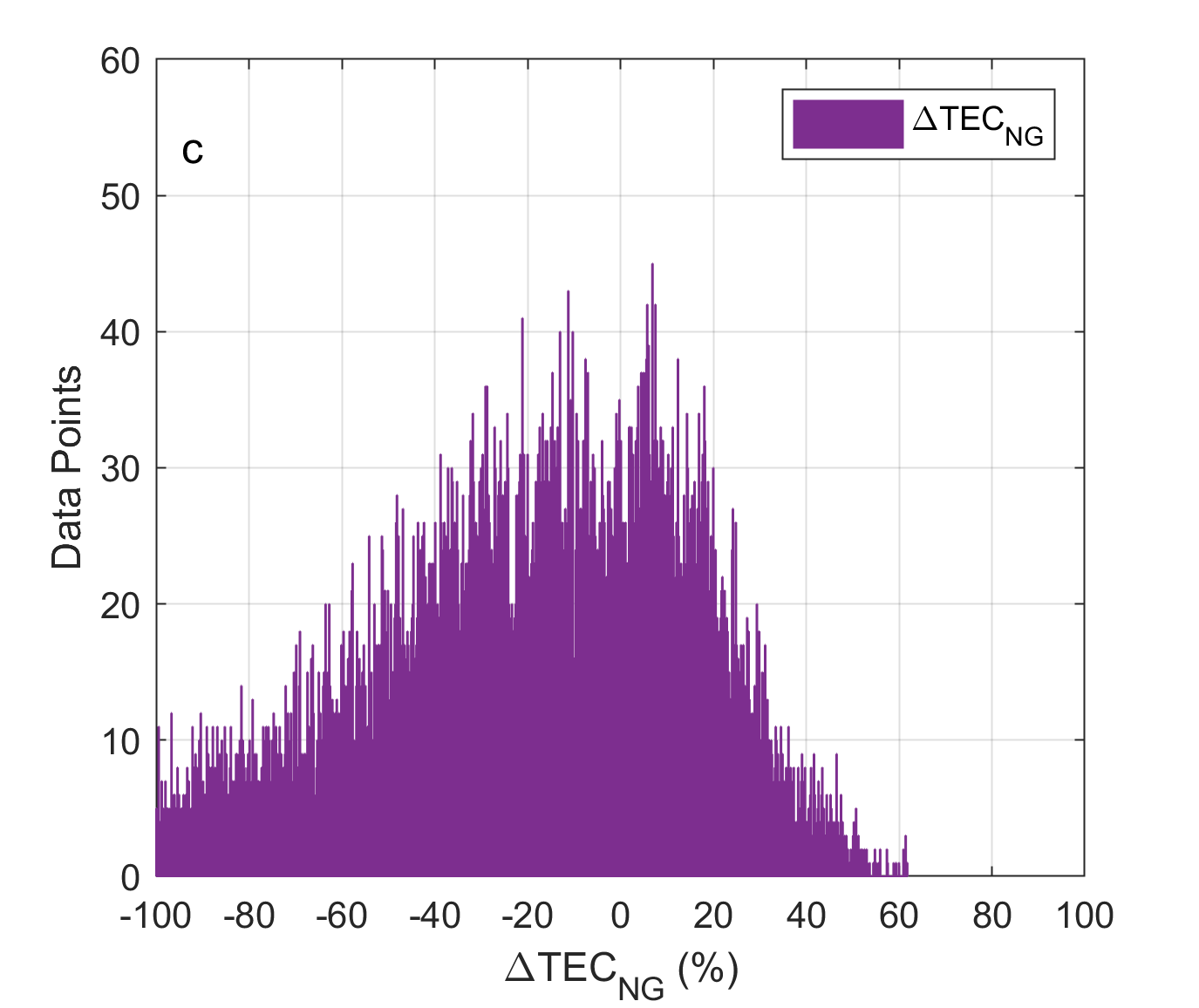}\\
\caption{$\Delta TEC_{NG}$(\%) distribution during Quiet days combining all NavIC satellites (a) without any corrections (b) after making diurnal minimum value corrections without removal of anomalous data (c) after making diurnal minimum value corrections with removal of anomalous data}
\end{figure}

\newpage
\begin{figure}
    \centering
\includegraphics[width=2in,height=2in]{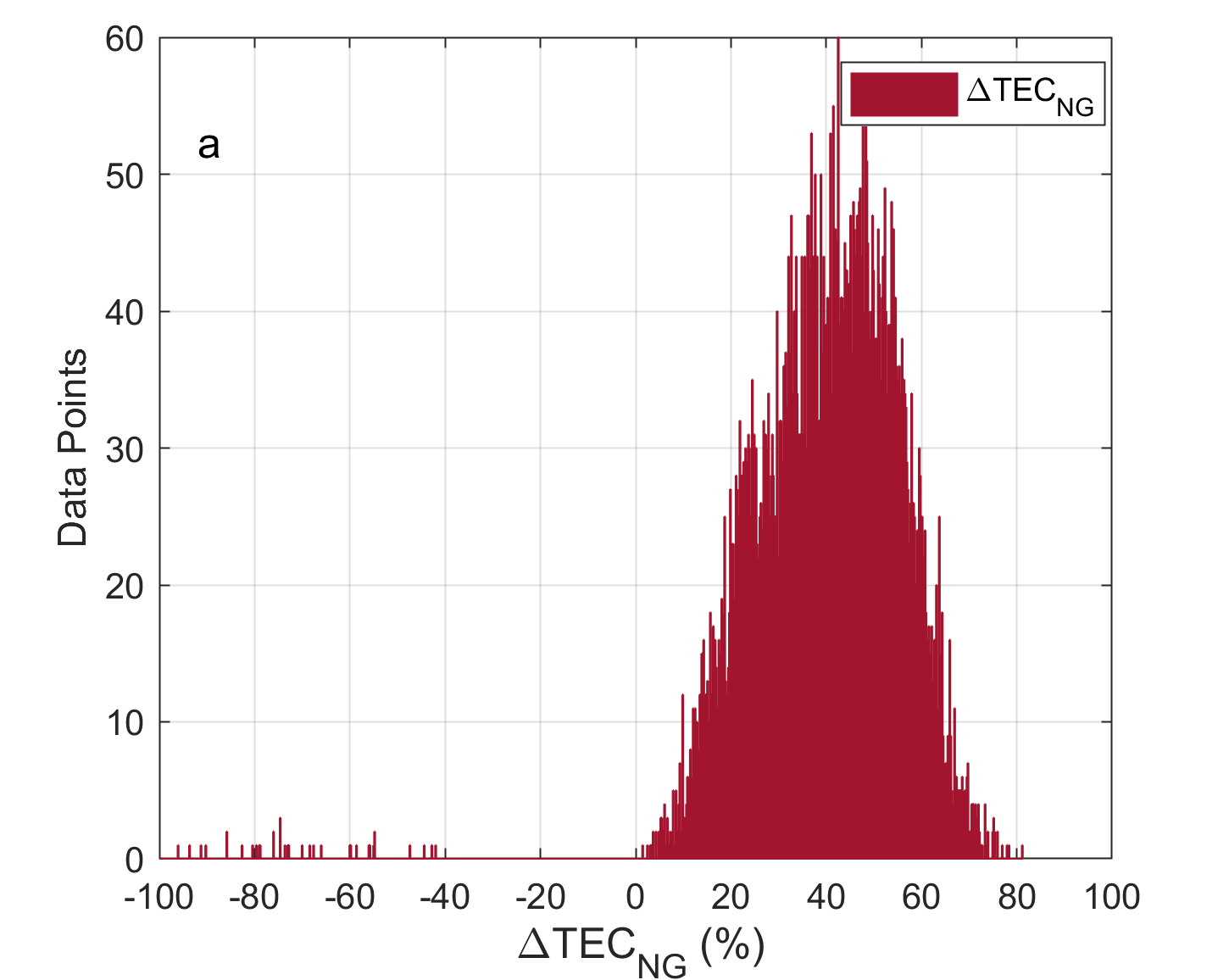}
\includegraphics[width=2in,height=2in]{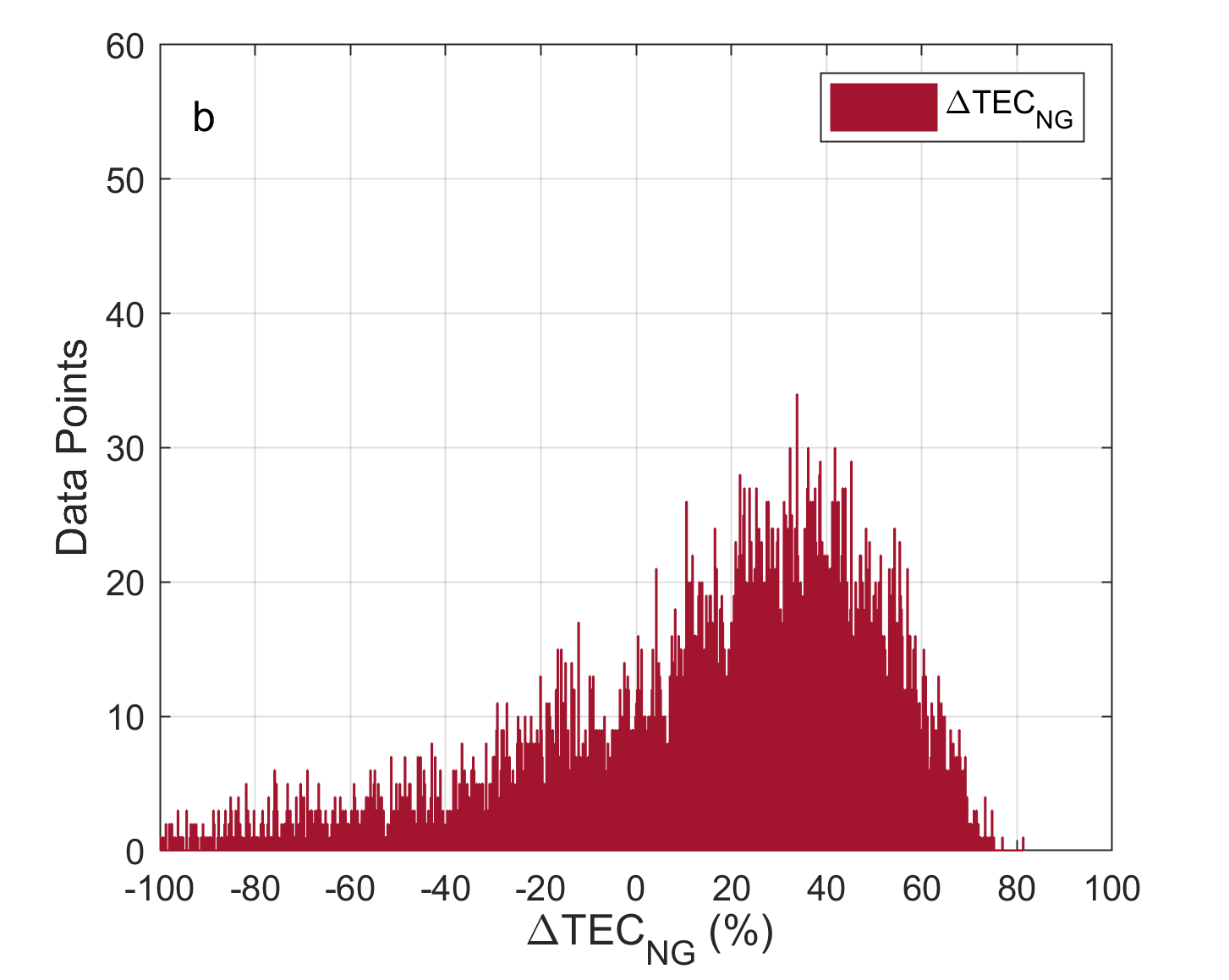}
\includegraphics[width=2in,height=2in]{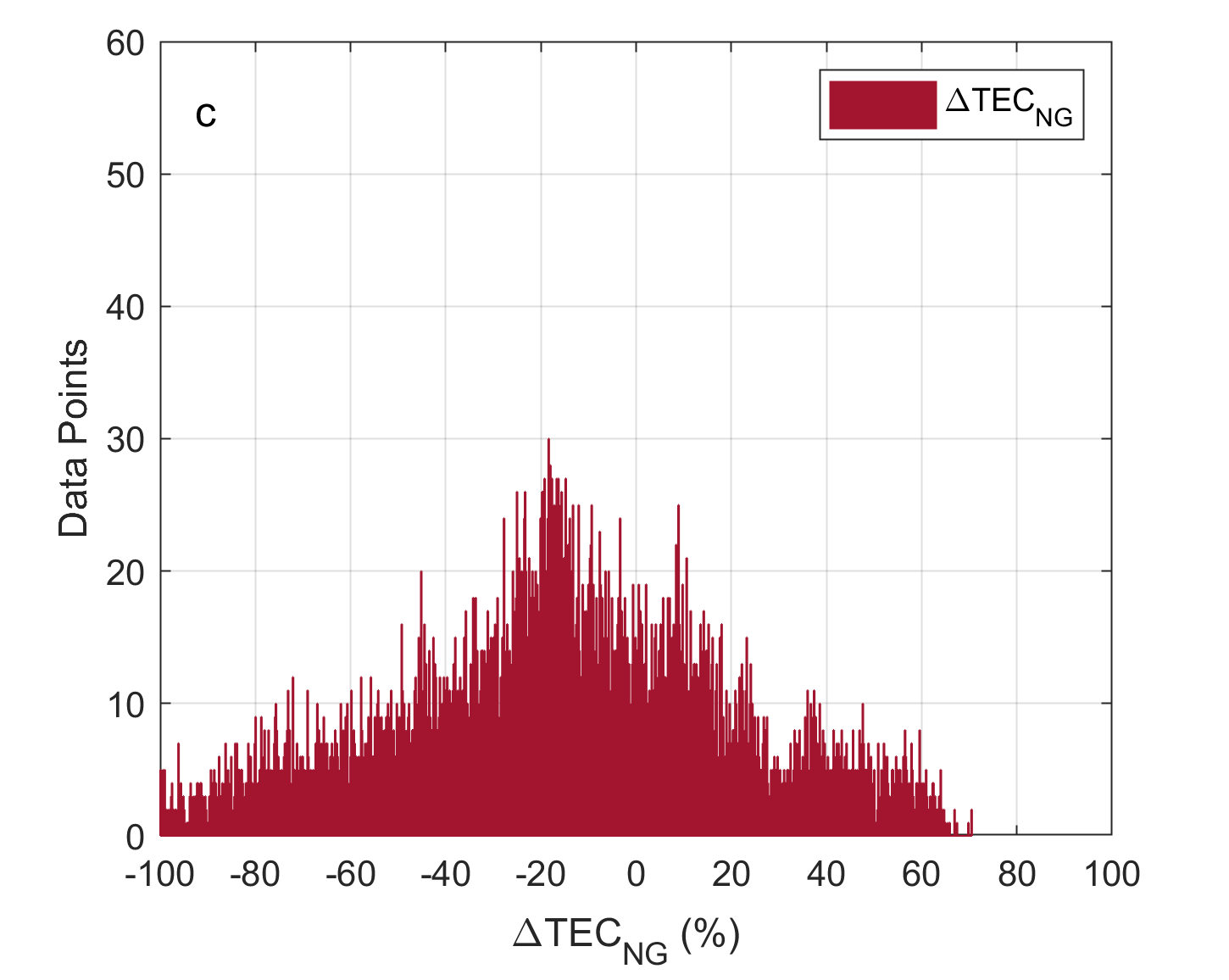}\\
\caption{$\Delta TEC_{NG}$(\%) distribution during Disturbed days combining all NavIC satellites (a) without any corrections (b) after making diurnal minimum value corrections without removal of anomalous data (c) after making diurnal minimum value corrections with removal of anomalous data}
\end{figure}

\newpage
\begin{figure}
\centering
\includegraphics[width=6in,height=4in]{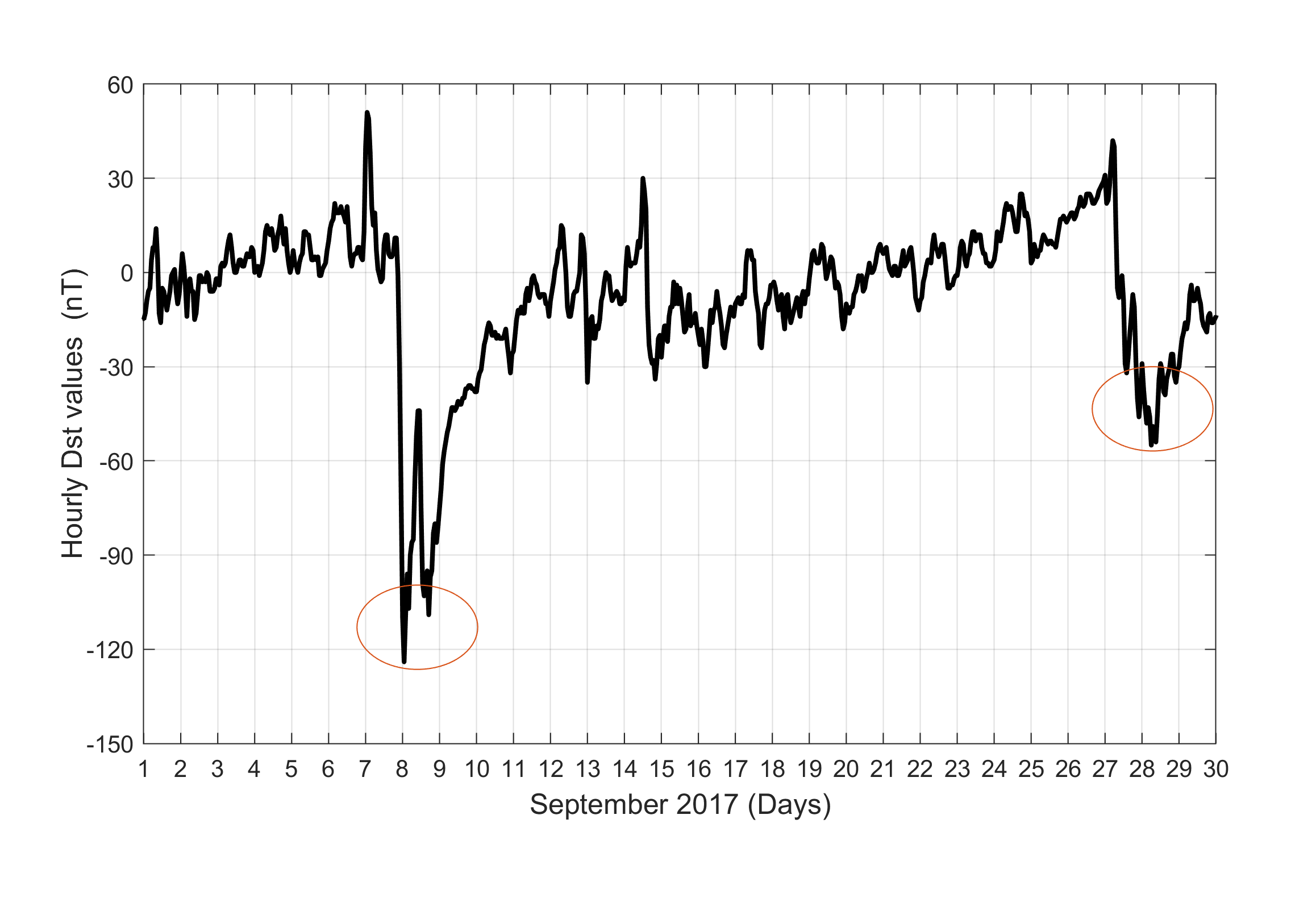}
\caption{Hourly Dst values (nT) plotted as a function of days of the month of September 2017. Instances of intense storm on September 8 and moderate storm on September 28 are observed as Dst reached a minimum (marked as red ellipses) of -124 nT on September 8 and -55 nT on September 28, 2017.}
\end{figure}

\newpage
\begin{figure}
\centering
\includegraphics[width=6in,height=4in]{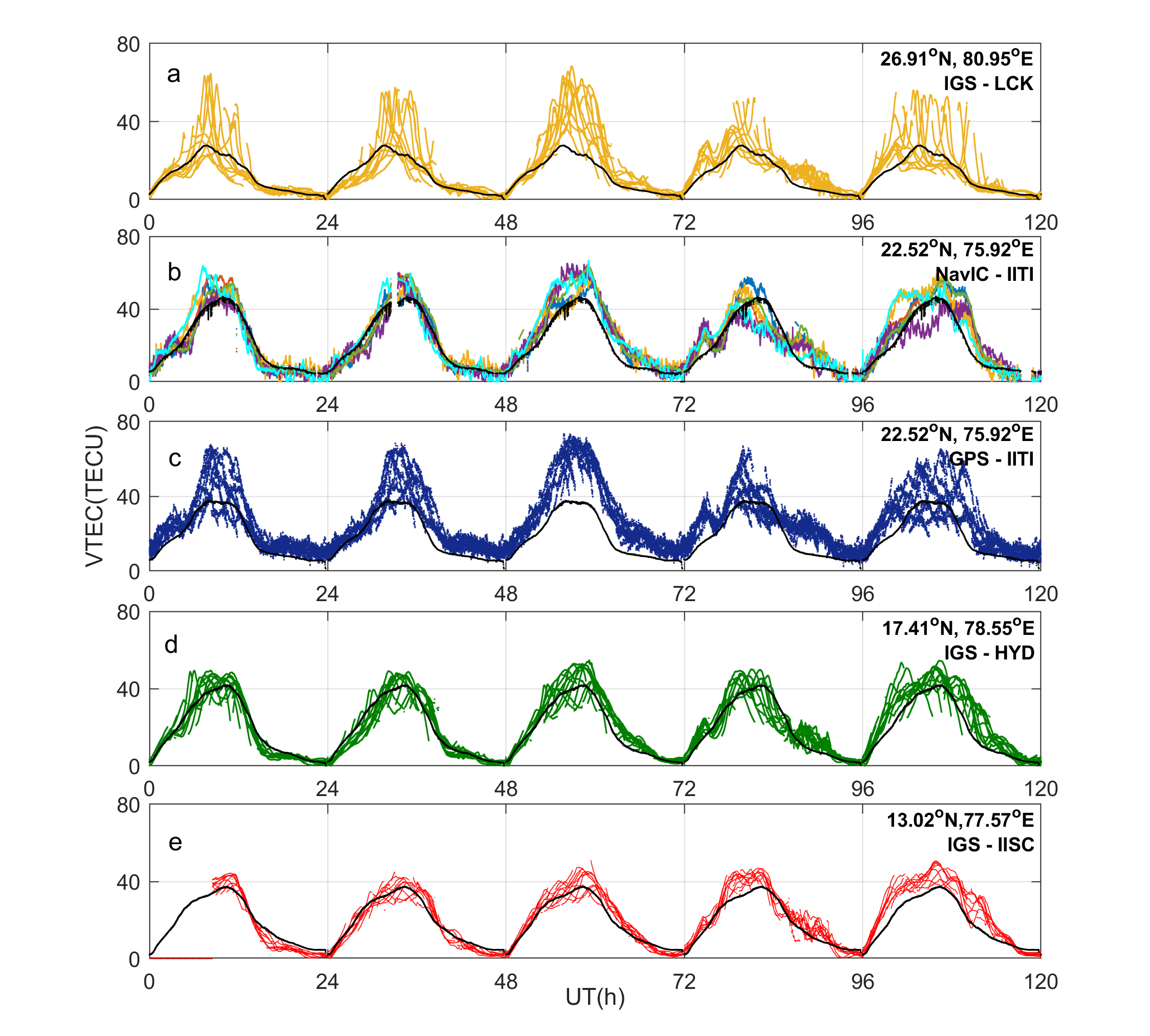}
\caption{Variation of VTEC (in TECU) as a function of UT (in hours) during period from September 5-9, 2017 for the stations: (a) Lucknow, (b-c) Indore, (d) Hyderabad and (e) Bengaluru. The monthly mean TEC values for the respective stations are shown in black.}
\end{figure}

\newpage
\begin{figure}
\centering
\includegraphics[width=6in,height=4in]{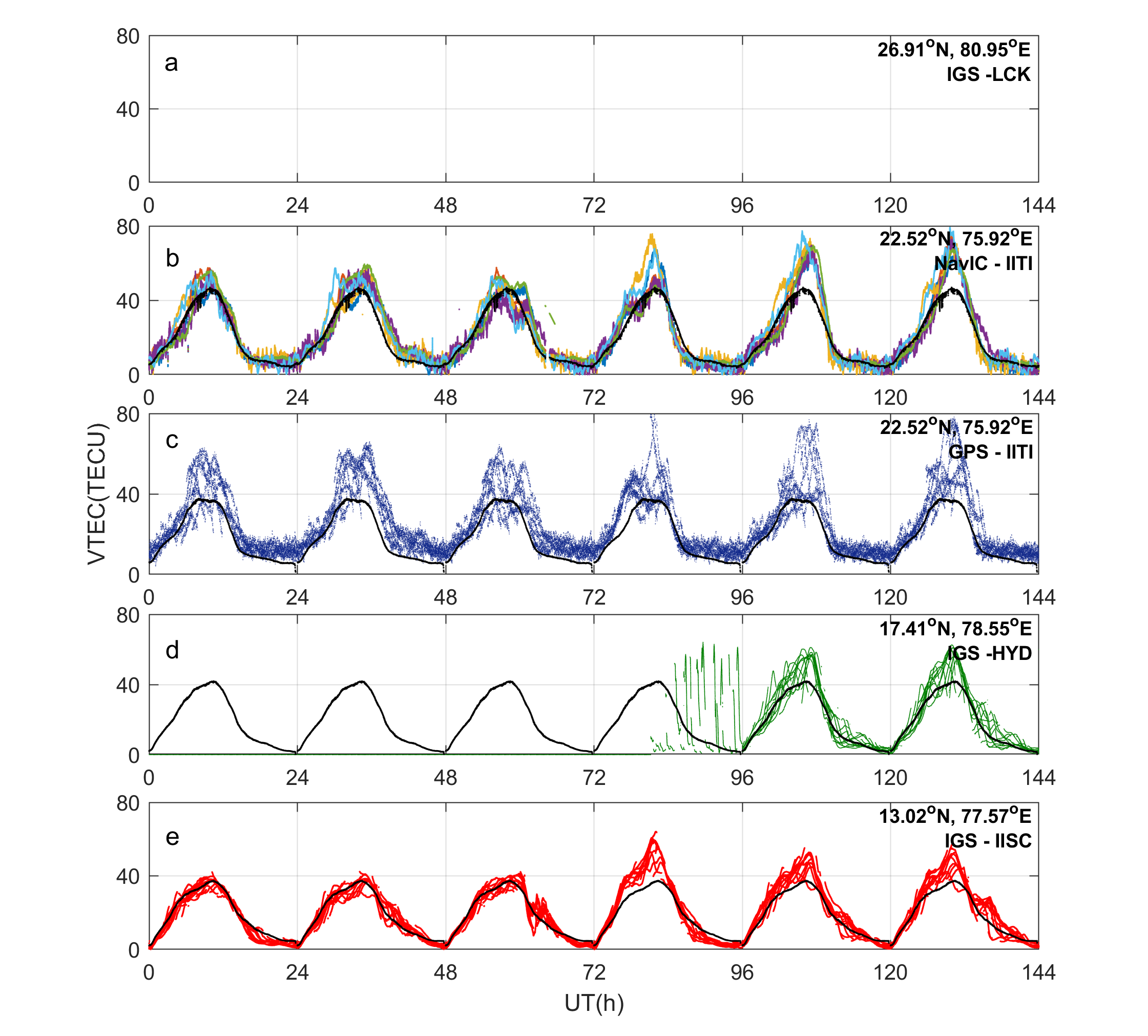}
\caption{Variation of VTEC (in TECU) as a function of UT (in hours) during period from September 25-30, 2017 for the stations:(a) Lucknow, (b-c) Indore, (d) Hyderabad and (e) Bengaluru. The monthly mean TEC values for the respective stations are shown in black.}
\end{figure}

\end{document}